\documentstyle[12pt]{article}
\newskip\AboveMaplePlot
\newskip\BelowMaplePlot
\newskip\AboveMapleSkip
\newskip\BelowMapleSkip
\newdimen\LeftMapleSkip
\newdimen\MaplePlotHeight
\newdimen\MaplePlotWidth
\newdimen\MapleSepLineWidth
\newdimen\MapleSepLineHeight \MapleSepLineHeight = 0.01in
\newif\ifMaple		\Maplefalse
\newif\ifMapleFirstLine
\newif\ifMaplePrompt
\newcount\MapleTab	\MapleTab = 8
\newcount\MapleParCount   \MapleParCount=0
\newtoks\MaplePromptString \MaplePromptString = {}
\MaplePromptString = {\raise 1pt \hbox{$\scriptstyle>$\space}}
\BelowMapleSkip = \AboveMapleSkip
\let\MapleSepLineWidth\linewidth  
\let\MapleFont\tt
\let\MapleSize\small
\MapleFirstLinefalse  
\MaplePrompttrue      
\makeatletter

\def\MakeActive#1{\catcode `#1 = \active\relax }
\def\MakeTabActive{\MakeActive{\^^I}}
\def\MakeEolActive{\MakeActive{\^^M}}
\newif\if@IgnoreNewLine
\@IgnoreNewLinetrue
\def\@MaplePar{\nopagebreak[3]\par\@@par}%
\def\MapleSpace{\ }
\def\@ObeySpaces{\MakeActive{\ }\@@ObeySpaces\relax}
{%
\MakeActive{\ }\gdef\@@ObeySpaces{\edef {\MapleSpace}}%
}%
\newdimen\@MapleTabSize
\def\@ObeyTabs{\MakeTabActive\@@ObeyTabs\relax}
{%
\MakeTabActive\gdef\@@ObeyTabs{\def^^I{\@MapleTab}}
}%
\def\@MapleTab{%
  \leavevmode   
  \egroup      
  \dimen0=\wd0 
  \divide\dimen0 by \@MapleTabSize
  \advance\dimen0 by 1sp
  \multiply \dimen0 by \@MapleTabSize
  \wd0 = \dimen0
  \box0
  \setbox0 = \hbox\bgroup
}
\def\@ObeyEol{\MakeEolActive\@@ObeyEol\relax}%
{%
  \MakeEolActive %
  \gdef\@@ObeyEol{%
    \let^^M=\@MapleEol%
  }%
}
\def\@MapleEol{%
  \if@IgnoreNewLine
  \else
    \leavevmode%
    \egroup%
    \box 0%
    \@MaplePar%
  \fi
  \@IgnoreNewLinefalse
}
\def\@SetupMapleTty#1{%
\begingroup
  \setbox 0 = \hbox{\MapleSize\MapleFont X}
  \@MapleTabSize = \wd0
  \multiply\@MapleTabSize by \MapleTab
  \setbox0 = \hbox{\relax}  
  \rightskip = 0pt
  \parindent=0pt
  \parskip = 0pt
  \leftskip=\LeftMapleSkip
  \parfillskip = 0pt plus 1fil
  \spaceskip = 0pt   \xspaceskip = 0pt
\ifnum #1 = 0 \@IgnoreNewLinetrue\else\@IgnoreNewLinefalse\fi%
\everypar = {\EveryParZ}%
\def\EveryParZ{%
  \ifMaplePrompt
    \the\MaplePromptString
  \fi
    \setbox0 = \hbox \bgroup
}%
\MapleSize\MapleFont%
\lineskiplimit=0\normalbaselineskip
\baselineskip=0.8\normalbaselineskip
\@noligs%
\let\do\@makeother \dospecials
\catcode ``=\active
\frenchspacing%
\@ObeySpaces%
\@ObeyTabs%
\@ObeyEol%
}
\def\@EndMapleInput{%
  \endgroup  
}
\let\@EndMapleTtyout\@EndMapleInput
\begingroup
  \catcode`| = 0  
  \catcode`( = 1  
  \catcode`) = 2  
  \catcode`@ = 11 
  \catcode`\{ = 12 
  \catcode`\} = 12 
  \catcode`\\ = 12 
  |gdef|@BeginMapleInput#1\end{mapleinput}(#1|end(mapleinput))%
  |gdef|@BeginMapleTtyout#1\end{maplettyout}(#1|end(maplettyout))%
|endgroup
\def\mapleinput{
\@MaplePar%
\if@minipage\removelastskip\vskip-1.3ex
\else\vskip\AboveMapleSkip\fi%
\@SetupMapleTty{0}
\@BeginMapleInput
}
\def\endmapleinput{
  \@EndMapleInput
  \nopagebreak[3]
}
\def\maplettyout{
  \MaplePromptfalse	  
  \nopagebreak[3]\@MaplePar
  \removelastskip
  \@SetupMapleTty{0}   
  \@BeginMapleTtyout
}
\def\endmaplettyout{
  \@EndMapleTtyout
  \vskip\BelowMapleSkip
  \pagebreak[3]
  \par
}
\newenvironment{maplelatex}{
  \parindent=0pt
  \parskip = 0pt
  \parfillskip = 0pt plus 1fil
  \lineskiplimit=0\normalbaselineskip
  \baselineskip=0.6\normalbaselineskip
  \abovedisplayskip=.4\AboveMapleSkip
  \belowdisplayskip=.4\AboveMapleSkip
  
  \removelastskip
  \nopagebreak[3]\@@par
}{%
\baselineskip=\normalbaselineskip%
\pagebreak[3]\relax}
\newcommand{\maplesepline}{\vskip \parskip%
\hrule\@height\MapleSepLineHeight\@width\MapleSepLineWidth%
\vskip \parskip\relax}
\makeatother
\evensidemargin=\oddsidemargin
\MaplePrompttrue      
\MaplePromptString = {\raise 1pt \hbox{$\scriptstyle>$\space}}
\BelowMapleSkip = \AboveMapleSkip
\let\MapleSepLineWidth\linewidth  
\let\MapleFont\tt     
\let\MapleSize\small  
\MapleFirstLinefalse  
\title{Junctions and thin shells in general relativity \\
 using computer algebra \\
I: The Darmois-Israel Formalism}

\author{Peter Musgrave\dag
        and Kayll Lake\ddag \\
   Dept. of Physics, Queen's University \\
   Kingston, Ontario, Canada }

\date{December 1, 1995}

\begin{document}

\maketitle

\abstract{We present the {\tt GRjunction} package which allows
boundary surfaces and thin-shells in general relativity to be
studied with a computer algebra system. Implementing the
Darmois-Israel thin shell formalism requires a careful
selection of definitions and algorithms to ensure that
results are generated in a straight-forward way. We have used the
package to correctly reproduce a wide variety of examples
from the literature. We present several of these verifications
as a means of
demonstrating  the packages capabilities. We then use
{\tt GRjunction} to perform a new calculation - joining two Kerr
solutions with differing masses and angular momenta along a
thin shell in the slow rotation limit.
}

\noindent \dag  Electronic address: musgrave@astro.queensu.ca \\
\noindent \ddag Electronic address: lake@astro.queensu.ca \\

\newpage

\section{Introduction}
The Darmois-Israel junction/thin-shell formalism
has found wide
application in general relativity and cosmology \cite{Darmois,Israel66}.
The junction of dust to Schwarzschild by Oppenheimer and Snyder
allowed the first insights into the nature of gravitational
collapse to a black hole \cite{OppSnyder}. Since Israel's landmark paper
\cite{Israel66} the formalism has been applied in a number of
contexts ranging from further studies of gravitational
collapse to the evolution of bubbles and domain walls
in a cosmological setting. \\

In this paper we describe the {\tt GRjunction} package we
have developed to assist relativists in the evaluation
of junction conditions and the parameters associated with
thin-shells (the package is available free of charge \cite{ftpSite}).
At the present time the package deals only with
non-null surfaces - although efforts to extend this to
null shells are underway.
{\tt GRjunction} runs under the computer algebra
system Maple \cite{Maple} in conjunction with GRTensorII \cite{GRT}.
Our goal in creating the package was to ensure that
it could easily recover all the standard shell
results in the literature (the bulk of which assume
spherical symmetry) without biasing the package towards
spherical symmetry in any way - allowing users to probe
the relatively unstudied area of non-spherical
shells and junctions. The package is necessarily interactive
allowing users to manipulate results and determine
the conditions for junctions. \\

We begin by outlining the shell formalism to establish
notation and motivate choices of algorithms which
we describe in the following section. We then demonstrate the
package by repeating some standard junction and shell
calculations. Next we present some new results relating to
the study of shells around slowly spinning black holes.
To validate the package we re-executed a number of the
standard results in the literature. A summary of these
tests appears in the final section. \\

The intention of this paper is to describe the junction
package we have developed and not to review the vast
literature on junctions and thin shells.
The references we have chosen are not always the first or
simplest treatment of a problem and in some cases we have
deliberately selected examples which differ from the standard
treatments to test the robustness of our package. \\

\section{The Formalism}
In this section we review the junction formalism to
establish notation. For more detailed discussions see
e.g.\cite{MTW}-\cite{juncRevLast}.\\

Consider two spacetimes (Lorentzian manifolds with signature
$(+ + + -)$) $M^+$ and $M^-$ with metrics
$g^+_{\alpha \beta}(x_+^\gamma)$ and
$g^-_{\alpha \beta}(x_-^\gamma)$ in the coordinate systems
$x_+^\gamma$ and $x_-^\gamma$. Within these
spacetimes define two non-null 3-surfaces $\Sigma^+$ and
$\Sigma^-$ (in $M^+$ and $M^-$ respectively) with metrics
$g^+_{ij}(\xi_+^c)$ and
$g^-_{ij}(\xi_-^c)$ in the coordinates $\xi_+^c$ and
$\xi_-^c$ which decompose each of the 4-manifolds into
two distinct parts.
(Greek indices range over the coordinates of the 4-manifold and
Roman indices over the coordinates of the 3-surfaces).
We label the distinct parts of $M^+$ created
by $\Sigma^+$ as $M_1^+$ and $M_2^+$
and likewise for $M^-$. The junction/shell formalism constructs
a new manifold $\cal M$ by joining one of the distinct parts
of $M^+$ to one of the distinct parts of $M^-$
by the identification $\Sigma^+ = \Sigma^- \equiv \Sigma$.
Clearly there are four possibilities, i.e.
$M_1^+ \cup M_1^-, M_2^+ \cup M_1^-, M_1^+ \cup M_2^-,
M_2^+ \cup M_2^-$. The assumed isometry between the points
on the surface is often (but need not always be) the identification
$\xi_+^c = \xi_-^c$. \\

What now follows holds simultaneously for $M^+$ and $M^-$ and so
we drop the $\pm$ distinction in this paragraph. The parametric
equation for $\Sigma$ is of the form
\begin{equation}
\label{eqn-surf}
f(x^\alpha(\xi^a)) = 0.
\end{equation}
We assume that $\Sigma$ is non-null.
The unit 4-normals to $\Sigma$ in $M$ are given by
\begin{equation}
\label{eqn-ndef}
n_\alpha =\pm \frac{1}{(\mid g^{\beta \gamma}
                  \frac{ \partial f}{\partial x^\beta}
                  \frac{ \partial f}{\partial x^\gamma} \mid )^{1/2} }
               \frac{\partial f}{\partial x^\alpha}.
\end{equation}
We assume $n_\alpha \neq 0$ and label $\Sigma$ as timelike (spacelike)
for $\Delta \equiv - n_\alpha n^\alpha = -1 (1)$. The three basis vectors
tangent to $\Sigma$ are
\begin{equation}
e^\alpha_{(a)} = \frac{\partial x^\alpha}{\partial \xi^a}
\end{equation}
which give the induced metric on $\Sigma$ by
\begin{equation}
g_{i j} = \frac{\partial x^\alpha}{\partial \xi^i}
          \frac{\partial x^\beta} {\partial \xi^j} g_{\alpha \beta}.
\end{equation}
The extrinsic curvature (second fundamental form) is given by
\begin{eqnarray}
\label{eqn-Kij1}
K_{i j} & = & \frac{\partial x^\alpha}{\partial \xi^i}
              \frac{\partial x^\beta}{\partial \xi^j} \nabla_\alpha
n_\beta          \\
\label{eqn-Kij2}
& = & -n_\gamma \left(
\frac{\partial^2 x^\gamma}{\partial \xi^i \partial \xi^j}
+ {\Gamma^\gamma}_{\alpha \beta}
  \frac{\partial x^\alpha}{\partial \xi^i}
  \frac{\partial x^\beta}{\partial \xi^j}
  \right).
\end{eqnarray}
We define $u^\alpha$ the four tangent to $\Sigma$ and $\dot{u}^\alpha
(\equiv u^\beta \nabla_\beta u^\alpha)$ the four-acceleration.
Combining these with the above relations gives
\begin{equation}
\label{eqn-jumpnu}
\left[ n_\alpha \dot{u}^\alpha \right] = - u^i u^j \left[ K_{ij} \right] \\
\end{equation}
and
\begin{equation}
\label{eqn-meannu}
\overline{n_\alpha \dot{u}^\alpha} =  - u^i u^j \overline{K_{ij}}
\end{equation}
where $\left[ X \right] \equiv X^+\mid_{\Sigma} - X^-\mid_{\Sigma}$
and $\overline{X} \equiv (X^+ \mid_\Sigma + X^- \mid_\Sigma)/2$
with $X^\pm\mid_{\Sigma}$ denoting the limiting values of $X$ on $\Sigma$.

The Darmois conditions for the joining of a part of $M^+$ to a part of
$M^-$ are
\begin{equation}
\label{eqn-jumpg}
\left[ g_{ij} \right] = 0
\end{equation}
and
\begin{equation}
\label{eqn-jumpK}
\left[ K_{ij} \right] = 0
\end{equation}

If both (\ref{eqn-jumpg}) and (\ref{eqn-jumpK}) are satisfied we refer to
$\Sigma$ as a boundary surface. If only (\ref{eqn-jumpg}) is
satisfied then we refer to $\Sigma$ as a thin-shell. \\

Conditions  (\ref{eqn-jumpg}) and (\ref{eqn-jumpK})
require a common coordinate system on $\Sigma$ and this
is easily done if one can set $\xi_+^a = \xi_-^a$. Failing this,
establishing (\ref{eqn-jumpg}) requires a solution to the three dimensional
metric equivalence problem.
Condition (\ref{eqn-jumpK}) as it stands is ambiguous since the
orientation of the 4-vector field
$n_\alpha \equiv n^\pm_\alpha \mid_{\Sigma}$ has not been specified.
The Israel formalism requires the normals in $\cal M$
to point from $M_A^-$ to $M_B^+$ (where A denotes the part of $M^-$
and B denotes the part of $M^+$ we wish to use to form $\cal M$).
Clearly the sign of the normal vectors are crucial since e.g.
$n^-_\alpha$ points away from the portion of $M^-$ which will be
used in forming $\cal M$. Hence an
understanding of which side  $n^-_\alpha$ points into is key. In general
this can be done by considering a trajectory in $M^-$ through $\Sigma^-$
with tangent $n^-_\alpha$ (and likewise for $n^+_\alpha$).
The majority of the existing literature deals with
spherical symmetry where the direction of the normal is clear, but
in more complicated examples (see below) great care must be taken.
Note that while there are two normal vectors in each of $M^\pm$ once
we have identified $\Sigma^-$ and $\Sigma^+$ there is a single
unique normal field to $\Sigma$ in $\cal M$. There may be
circumstances in which one can determine the differential relation
between the coordinates of $M^-_A$ and $M^+_B$ via
\begin{equation}
\label{eqn-nxform}
n^-_\alpha = \frac{\partial x_+^\alpha}{\partial x_-^\beta} n^+_\beta
\end{equation}
in an open neigbourhood of $\Sigma$ in $\cal M$ and this will give the
direction of one normal vector relative to the other.
(In the case of a boundary surface often only the sign of $n_\alpha$
on one side of $\Sigma$ needs to be determined.
Since $\nabla_\alpha n^\alpha = K^i_i$ in the case of a boundary surface
(\ref{eqn-jumpK}) gives the useful relation
$\left[ \nabla_\alpha n^\alpha \right] = 0$.) \\

Some studies have left the sign of the normal vectors unspecified
to exhaustively study the taxonomies of all possible combinations
of $M^-$ and $M^+$ (usually excluding those which require shells
which violate energy conditions e.g. \cite{Sato,Maeda}).
The junction
package allows the user to leave the sign unspecified but
we take the view that
the ``typical'' starting point is to explicitly choose the signs of
the normal vectors so that a particular combination of
$M^-_A$ and $M^+_B$ can be studied. \\

Once we have selected signs in (\ref{eqn-ndef}) there is no
ambiguity in (\ref{eqn-jumpK}) and we use
(\ref{eqn-jumpg}) and (\ref{eqn-jumpK}) in conjunction with the Einstein tensor
$G_{\alpha \beta}$ to obtain the identities
\begin{equation}
\label{eqn-Gnn}
\left[ G_{\alpha \beta} n^\alpha n^\beta \right] = 0
\end{equation}
and
\begin{equation}
\label{eqn-Gfn}
\left[ G_{\alpha \beta}
   \frac{\partial x^\alpha}{\partial \xi^i}
   n^\beta \right] = 0.
\end{equation}
(Note that (\ref{eqn-Gnn}) and (\ref{eqn-Gfn}) do not guarantee
(\ref{eqn-jumpg}) and (\ref{eqn-jumpK}).)
This shows, for example, that for timelike $\Sigma$ the flux
through $\Sigma$ (as measured comoving with $\Sigma$)
is continuous.  \\

The Israel formulation of thin shells follows from the
Lanczos equation
\begin{equation}
\label{eqn-Lanc}
S_{ij} = \frac{\Delta}{8 \pi}(\left[K_{ij} \right] - g_{ij} \left[{K_i}^i
\right])
\end{equation}
and we refer to $S_{ij}$ as the surface stress-energy tensor of
$\Sigma$. The ``ADM'' constraint
\begin{equation}
\nabla_j K^j_i - \nabla_i K  =
    G_{\alpha \beta} \frac{\partial x^\alpha}{\partial \xi^i} n^\beta
\end{equation}
along with Einstein's equations then gives the conservation {\em identity}
\begin{equation}
\label{eqn-claw}
\Delta \nabla_i S^i_j =
  \left[T_{\alpha \beta} \frac{\partial x^\alpha}{\partial \xi^i}
  n^\beta \right].
\end{equation}
The ``Hamiltonian'' constraint
\begin{equation}
G_{\alpha \beta} n^\alpha n^\beta =
  (\Delta (^3 R) + K^2 - K_{ij} K^{ij})/2
\end{equation}
gives the evolution {\em identity}
\begin{equation}
\label{eqn-elaw}
-S^{i j} \overline{K}_{i j} = \left[ T_{\alpha \beta} n^\alpha n^\beta \right].
\end{equation}

The identities (\ref{eqn-claw}) and (\ref{eqn-elaw}) do not give information
about the dynamics of the shell. The evolution of the shell
stems from a phenomenological interpretation and the Lanczos equation
(\ref{eqn-Lanc}). \\

\subsection{Phenomenology}
The standard phenomenology associated with $\Sigma$ is introduced
as follows: let $\xi^a$ be the coordinates intrinsic to $\Sigma$ and
consider a trajectory $\xi^a(\tau)$ with associated 3-tangent
\begin{equation}
u^a = \frac{d\xi^a}{d\tau}
\end{equation}
where $u^a u_a = \Delta$. For a timelike surface we view the curve
as the worldline of a (possibly hypothetical) particle in the
surface with $\tau$ the proper time. (If the surface is spacelike
the phenomenology is a formal analogy).
The associated four tangent is
\begin{equation}
u^\alpha_\pm = \frac{\partial x^\alpha_\pm}{\partial \xi^a} u^a =
\frac{dx^\alpha_\pm}{d \tau}.
\end{equation}
Equivalently
\begin{equation}
u_a = u^\pm_\alpha \frac{\partial x^\alpha_\pm}{\partial \xi^a}
\end{equation}
We view $\Sigma$ as covered by a 3-vector field $u^a$.
Note that $\tau$ is defined curve by curve on $\Sigma$ but not in
general over the entire surface. If $\tau$ is
defined over the entire surface then we label this as a
``one-parameter surface''.
The surface energy density
$\sigma$ associated with the trajectory $u^a$ on $\Sigma$ is
defined by
\begin{equation}
S_{ab} u^b = - \sigma(\xi^c) u_a - q_a
\end{equation}
where $u_a q^a = 0$. The inclusion of $q_a$ represents an intrinsic
energy flux orthogonal to the prescribed intrinsic velocity field.
The surface energy density $\sigma(\xi^c)$ is not in
general an eigenvalue but is given by
\begin{equation}
\label{eqn-sigma}
\sigma(\xi^c) = - \Delta S_{ab} u^a u^b.
\end{equation}

In analogy to a perfect 4-fluid we suppose that $q_a = 0$ and that
$S_{ab}$ takes the form
\begin{equation}
S_{ab} = - \Delta (\sigma(\xi^c) + p(\xi^c))u_a u_b + p(\xi^c) g_{ab}.
\end{equation}
This defines the surface pressure ( - surface tension)
\begin{equation}
\label{eqn-pressure}
p(\xi^c) = ( \sigma(\xi^c) + S^a_a)/2.
\end{equation}
It now follows from (\ref{eqn-claw}) that
\begin{equation}
\label{eqn-conservation}
\dot{\sigma} + (\sigma + p)\Phi = - \Delta
  \left[T_{\alpha \beta} \frac{\partial x^\alpha}{\partial \xi^i}
  n^\beta \right]
\end{equation}
where $\dot{~}$ signifies the intrinsic 3-derivative along
$u^a$ ( $\dot{\sigma} = u^a \nabla_a \sigma$) and $\Phi$ is the
three-expansion of $u^a$ ($\Phi \equiv \nabla_a u^a$).
{}From (\ref{eqn-Kij1}), the Lanczos equation (\ref{eqn-Lanc}) and the
phenomenological equations (\ref{eqn-sigma}) and (\ref{eqn-pressure}) we have
\begin{equation}
\left[ n_\alpha \dot{u}^\alpha \right] = 8 \pi ( p + \sigma /2).
\end{equation}
{}From (\ref{eqn-Kij2}), (\ref{eqn-Lanc}) and the identity (\ref{eqn-elaw})
the phenomenology gives
\begin{equation}
\label{eqn-lawN}
\Delta (\sigma + p) \overline{n_\alpha \dot{u}^\alpha} + p \overline{K^i_i}
= - \Delta \left[ T_{\alpha \beta} n^\alpha n^\beta \right].
\end{equation}

In spherical spacetimes the first integral of
the evolution equation (\ref{eqn-lawN})
is given by the identity \cite{LakeIdentity}
\begin{equation}
\label{eqn-lakeid}
\dot{R}^2 = \Delta + \left( \frac{\left[ m \right]}{M} \right)^2
   - \frac{2 \Delta \overline{m}}{R} +
   \left( \frac{M}{2R} \right)^2
\end{equation}
where $R = R(\tau), \dot{~} \equiv d/d\tau$ and $M = 4 \pi \sigma R^2$.
The effective mass, $m_\pm$ is given by
\begin{equation}
m_\pm \equiv \frac{1}{2} (^{(4)}g^\pm_{\theta
\theta})^{3/2}~^{(4)}{R^\pm_{\theta \phi}}^{~~\theta \phi}.
\end{equation}

\section{Computer Implementation}
Specifying the junction formalism for a computer
algebra system requires a careful choice of specific object
definitions and recognition of several standard calculus manipulations.
Joining spacetimes requires the simultaneous consideration
of four metrics, so the underlying general relativity
software must permit this. In order to make the package
reproduce the existing literature on spherical
shells special consideration must be given to the choice of
object definitions for one-parameter shells. We first present an
overview of the package and then
discuss the choices we have made in implementing the
junction/thin shell package for the Maple
version of GRTensorII. \\

\subsection{Overview}
The {\tt GRjunction} package provides the user with a means to specify a
surface,
calculate intrinsic and extrinsic quantities on the surface,
identify two
such surfaces and evaluate whether a boundary surface or thin-shell results.
The specification of a surface is done by invoking the command
{\tt surf} which then prompts the user for the necessary information.
Once a surface has been specified objects defined on the surface
(e.g. $K_{ij}$) can be calculated. The identification of two surfaces
is performed via the command {\tt join} which calculates
$\left[ g_{ij} \right]$ and displays the result.
If $\left[ g_{ij} \right] \ne 0$ the user can
manipulate this expression in an attempt to
determine restrictions on metric functions
which will ensure $\left[ g_{ij} \right] = 0$ (see the first example
below).
The jump or
mean of any quantity in the joined manifold can be evaluated by
means of the operators {\tt Jump} and {\tt Mean}. Hence to
determine $\left[ K_{ij} \right]$ the user would refer to
the object \verb|Jump[K(dn,dn)]|; to determine $\overline{K^{ij}}$,
\verb|Mean[K(up,up)]|. Standard quantities and
equations for the joined spacetimes can be calculated in
a straight-forward manner. For example to determine $S_i^j$ the
user refers to the object \verb|S3(dn,up)|. (By convention we
list {\tt dn} indices ahead of {\tt up} indices for mixed two
index objects). \\

While we have emphasised the Darmois-Israel formalism, {\tt GRjunction}
can trivially evaluate the Lichnerowicz junction conditions
\cite{Lich} which consist of
\begin{eqnarray}
\label{eqn-Lich1}
\left[ g_{\alpha \beta} \right] & = & 0 \\
\label{eqn-Lich}
\left[ \partial g_{\alpha \beta}/ \partial x^\gamma \right] & = & 0
\end{eqnarray}
(Recall these conditions require admissible
coordinates and consequently are not as general the Darmois-Israel
conditions). The object $\partial g_{\alpha \beta}/ \partial x^\gamma$
is referred to as {\tt g(dn,dn,pdn)} ({\tt pdn} denoting the partial
derivative index) and so (\ref{eqn-Lich}) can be evaluated by
referring to the object \verb|Jump[g(dn,dn,pdn)]|. {\tt GRTensorII}
also allows users to work within the Newman-Penrose formalism and the
{\tt GRjunction} will allow users to evaluate jumps in the
spin coefficients across a surface. (This technique is employed in
e.g. \cite{Redmount}). \\

\subsection{The surface and related quantities}
In this section
we describe how to specify a non-null surface to the junction package
and the intrinsic (on $\Sigma$) and extrinsic quantities
which can be calculated with the package.
In this section we restrict attention to quantities which
are independent of phenomenology. \\

All the expressions in this section should implicitly
carry a $\pm$ designation (which we omit)
{\em with the exception of the} $\xi^i$,
the coordinates on $\Sigma$. For the package
to compare first and second
fundamental forms we must have $\xi^i_+ = \xi^i_-$
The package does not currently consider the three-metric equivalence
problem. \\

In general to specify a surface $\Sigma$ in a space
$M$ with coordinates $x^\alpha$ in sufficient detail
to allow calculation of the first and second fundamental
forms we require
\begin{itemize}
\item the coordinates $\xi^i$ on $\Sigma$
\item the coordinate definition of $\Sigma$:
$x^\alpha = x^\alpha(\xi^i)$.
\item the parametric definition of $\Sigma$: $f(x^\alpha)=0$
\item the choice of normal vector sign in (\ref{eqn-ndef})
\end{itemize}

The essential idea of the
Darmois/Israel formalism is to use intrinsic quantities
in the description of all objects of interest on $\Sigma$.
However many of the definitions of objects on the surface
are in terms of quantities in $M$.
For example
(\ref{eqn-Kij2}) uses the normal vector which involves partial derivatives
with respect to the coordinates of $M$ and Christoffel symbols of $M$.
This can frequently be resolved simply by substituting the
coordinate definition of the surface
(i.e. $x^\alpha = x^\alpha(\xi^i)$) but this is
not always desirable.
Consider a metric which has an arbitrary function $u(r,\theta)$ and
a definition of $\Sigma$ with coordinates
$\xi^a = (\tilde \theta, \tilde \phi, \tau)$
and which has a coordinate definition which includes $r=f(\xi^i)$
and $\theta = \tilde \theta$.
The Christoffel symbols used in defining $K_{ij}$ may contain
partial derivatives of $u$ with respect to both $r$ and $\theta$.
Substitution
of $\theta = \tilde \theta$ will merely change the variable, but
the $r$ substitution will result in
$\partial u(f(\xi^i),\tilde \theta) / \partial f(\xi^i)$.
This is more than
notationally ugly - it constitutes an error in Maple; you cannot take
a partial derivative with respect to a function. In these cases we make
use of the Maple operator {\tt D} (instead of the Maple procedure {\tt diff}).
For example the representation of the $r$ derivatives of $f(r)$ and
$\nu(r,\theta)$ on the surface are represented in Maple as
\begin{eqnarray}
\frac{d}{dr} f(r) \mid_{r=R(t)} & \rightarrow & D(f)(R(t)) \\
\frac{d}{dr} \nu(r,\theta) \mid_{r=R(t)} & \rightarrow &
  D\left[ 1 \right](\nu)(R(t), \theta) \\
\end{eqnarray}
(the $\left[ 1 \right]$ indicates that the derivative acts on the
first argument of the function $\nu$). It is essential that we make
use of this form of representing derivatives to ensure that their
time dependence is handled properly in subsequent calculations.
While we could use the {\tt D} derivative throughout we prefer to
minimize it's use since in the Maple output we find it preferable to
have expressions with $\partial f(r)/ \partial r$ instead of
\verb|D(f)(r)|. Consequently object defintions which take derivatives
and then evaluate values on the surface make use of code which
substitutes in the coordinate relations and ensures that the substitution
does not cause errors in derivatives. This code  then minimizes the number
of {\tt D} derivatives which appear by converting extraneous occurances
back into Maple's usual {\tt diff} derivative. \\

\subsection{One-Parameter Surfaces}
A large part of the existing shell literature has dealt with
timelike spherical 3-surfaces within spherical 4-manifolds.
The majority of these analyses define one of the $\xi^a$
to be the proper time on the surface (what we referred to
above as one-parameter shells).
We must take some care in choosing algorithms for
objects in these cases since some of the routine steps in a hand calculation
are best avoided in a computer algebra approach.
All of the algorithm issues arise in the specification of a surface
in one of $M^\pm$ so we limit our discussion to the specification
of one surface. \\

First we describe two standard timelike surfaces
we will make reference to in our discussion:
a static spherically symmetric surface and the dynamic counterpart
within a spherically symmetric 4-manifold.
In the 4-manifold we take coordinates $(r, \theta, \phi, t)$ and on
the 3-surface coordinates $(\tilde \theta, \tilde \phi, \tau)$.
For the static shell we use the coordinate definitions
\begin{equation}
r= R, \theta = \tilde \theta, \phi = \tilde \phi, t = T_s(\tau)
\end{equation}
and a surface equation $r-R = 0$. For the dynamic shell we use
\begin{equation}
r= R(\tau), \theta = \tilde \theta, \phi = \tilde \phi, t = T_d(\tau)
\end{equation}
and a surface equation $r-R(\tau) = 0$. \\

In both cases it is conventional to ensure that the coefficient of $d\tau^2$
on $\Sigma$ is $-1$ (hence $\tau$ is the proper time on the shell).
For this reason we express $t$ as a function of $\tau$
and then make use of the constraint $u_\alpha u^\alpha = -1$
to eliminate e.g. $\partial T_d(\tau)/ \partial \tau$ from the quantities
we calculate on $\Sigma$. This facility is built into the {\tt surf}
command. The user is asked if the surface has a parameter which
governs its evolution. If the user so indicates then the package
evaluates $u_\alpha u^\alpha$ and asks if there is a quantity
to be eliminated. An attempt is then made to
solve the tangent constraint so that this quantity can be eliminated.
The tangent constraint is then automatically applied during the
calculation of $n_\alpha$ and $K_{ij}$. \\

We must specify an algorithmic
approach to calculating quantities such as $K_{ij}$ and
$\dot{u}^\alpha$ which recover the results for one-parameter shells
in a straight-forward way.
Not all object definitions are
equivalent because some require steps which are obvious in a hand
calculation but are difficult for a computer algebra system.
One example of this is the calculation of $\dot{u^\alpha}$.
For a one parameter shell with $x^\alpha = x^\alpha(\tau)$
we would proceed with the definition:
\begin{eqnarray}
\label{eqn-udot}
u^\beta \nabla_\beta u^\alpha & = &
  u^\beta \left(
   \frac{\partial u^\alpha(\tau)}{\partial x^\beta} + \Gamma^\alpha_{\beta
\gamma}
  u^\gamma \right) \\
 & = & \frac{dx^\beta(\tau)}{d\tau} \frac{\partial u^\alpha(\tau)}{\partial
x^\beta}
    + \Gamma^\alpha_{\beta \gamma} u^\gamma u^\beta \nonumber \\
\label{eqn-udot2}
 & = & \frac{du^\alpha}{d\tau} +
      \Gamma^\alpha_{\beta \gamma} u^\gamma u^\beta
\end{eqnarray}
While in a hand calculation using the definition of a total
derivative to
get the last line is an obvious step this is not so in a computer
algebra system. Once specific components are used in (\ref{eqn-udot})
and partial derivatives are taken,
recognizing terms which can collected into
total derivatives with respect to $\tau$ becomes a problem in
pattern matching. To avoid such problems we require that the user
indicate the variable to be used for total derivatives and we use the
relation (\ref{eqn-udot2}) instead of (\ref{eqn-udot}). The same
issue arises in the calculation of $K_{ij}$ for the dynamic shell
discussed above if we use (\ref{eqn-Kij1}) so we use
the definition (\ref{eqn-Kij2}). This definition has the advantage that
it is expressed in terms of derivatives with respect to intrinsic quantities
(although we still need to evaluate the Christoffel symbols
of the 4-manifold on $\Sigma$). \\

Another computer algebra issue arises in the calculation
of the normal vector.
For one-parameter surfaces it is customary to write
the equation of the surface in terms
of a function of a parameter (e.g. $r-R(\tau)=0$). Yet the
covariant derivative in $M$ will require partial derivatives
with respect to the $x^\alpha$. The junction package requires
that the surface equation be in terms of the $x^\alpha$
(so for the dynamic shell example above we specify $r-R(t)$ instead of
$r-R(\tau)$).
In these cases the package then applies the chain rule
\begin{equation}
\label{eqn-normal}
\frac{dr}{dt} \rightarrow \frac{dR}{dT} \rightarrow
\frac{dR(t)}{d\tau} \frac{d\tau}{dt} \rightarrow
\frac{dR(\tau)/d\tau}{dT(\tau)/d\tau}.
\end{equation}
where we have made use of the relations $r=R(\tau)$ and $t=T(\tau)$ and
that $\dot{T} \neq 0$
If the surface equation is governed by a parameter other than
$\tau$ then the normal vector and $K_{ij}$ can be expressed in
terms of intrinsic quantities only if the function
dependence on the other $\xi^i$ can be given explicitly
in the surface equation (e.g
$r - r(t) cos \theta = 0$). If all that is known is e.g.
$r - R(t,\theta)$ then we cannot make use of (\ref{eqn-normal})
and $n_\alpha$ and $K_{ij}$ will be left in terms of the
partial derivative with respect to the coordinates of $M$.\\

With these definitions and constraints we now have available the
quantities: $n^\alpha, u^\alpha,
\dot{u}^\alpha, g_{ij}$ and $K_{ij}$ in terms of intrinsic quantities.
We require a few additional objects which are calculated in a
straight-forward way. A summary of all the objects which can
be calculated on a surface
by the junction package is provided in Tables 1 and 2. \\


\begin{table}
\begin{center}
\begin{tabular}{|l|l|l|l|}  \hline
Object &   Definition &  Name  & Comments \\ \hline

$x^\alpha(\xi^i)$ & $x^\alpha(\xi^i)$ & {\tt xform(up)} &
    coordinate definition of $\Sigma$ \\
$\Delta$ & $- n^\alpha n_\alpha$ & {\tt utype} &  \\
$f$ & $f(x^\alpha(\xi^i))$ & {\tt surface} & parametric surface equation \\
$n_\alpha$ & $\partial f / \partial x^\alpha$ (normalized) & {\tt n(dn)} &
  the normal to $\Sigma$ \\
$u^\alpha$ &  $dx^\alpha(\xi^a)/d\tau$ & {\tt u(up)} & four tangent to $\Sigma$
\\
 & & & (one parameter shells only) \\
$\tau$ &  & {\tt totalVar} & shell parameter (e.g. proper time) \\
 & & & (one parameter shells only) \\
$\dot{u^\alpha} $ & $d u^\alpha / d\tau + \Gamma^\alpha_{\beta \gamma} u^\beta
u^\gamma$ &
  {\tt udot(up)} & four acceleration of $\Sigma$ \\
 & & & (one parameter shells only) \\
$m$ & $\frac{1}{2} (^{(4)}g_{\theta \theta})^{3/2}~^{(4)}{R_{\theta
\phi}}^{~~\theta \phi}$
   & {\tt mass} & mass (for spherical \\
 & & & one parameter shells only) \\ \hline
\end{tabular}
\end{center}
\caption{Objects in $M^\pm$}
\end{table}

%
%
%
%
\begin{table}
\begin{center}
\begin{tabular}{|l|l|l|l|}  \hline
Object &   Definition &  Name  & Comments \\ \hline

$\xi^i$ & & {\tt x(up)} & coordinates on $\Sigma$ \\
$g_{i j}$ & $\frac{ \partial x^\alpha}{\partial \xi^i}
            \frac{\partial x^\beta}{\partial \xi^j} g_{\alpha \beta}$ &
  {\tt g(dn,dn)} & first fundamental form \\
$K_{i j}$ & $-n_\gamma \left( \frac{\partial^2 x^\gamma}{\partial \xi^i
        \partial \xi^j}
            + \Gamma^\gamma_{\alpha \beta} \frac{\partial x^\alpha}{\partial
\xi^i}
            \frac{\partial x^\beta}{\partial \xi^j} \right)$ &
    {\tt K(dn,dn)} & second fundamental form \\
$K^i_i$ &  & {\tt trK} & \\
$K_{ij} K^{ij}$ & & {\tt Ksq} & \\
 & $(^{(3)}R+ K^2 + K_{ij}K^{ij})/2 = G_{\alpha \beta} n^\alpha n^\beta$
    & {\tt HCGeqn} & Hamiltonian constraint \\
 & $(^{(3)}R+ K^2 + K_{ij}K^{ij})/2 = 8 \pi T_{\alpha \beta} n^\alpha n^\beta$
    & {\tt HCTeqn} & Hamiltonian constraint \\
 & $\nabla_i K - \nabla_j K^j_i =
    G_{\alpha \beta} \frac{\partial x^\alpha}{\partial \xi^i} n^\beta $
 & {\tt PCGeqn} & Momentum constraint \\
 & $\nabla_i K - \nabla_j K^j_i =
    8 \pi T_{\alpha \beta} \frac{\partial x^\alpha}{\partial \xi^i} n^\beta $
 & {\tt PCTeqn} & Momentum constraint \\
$u^i$ & $\frac{\partial x^\alpha}{\partial \xi^i} u_\alpha$ & {\tt u3(up)} &
three tangent \\
 & & & (one parameter surf. only) \\
$\Phi$ & $\nabla_i u^i$ & {\tt u3div} & three divergence \\
 & & & (one parameter surf. only) \\
 & & & \\ \hline
\end{tabular}
\end{center}
\caption{Objects on $\Sigma^\pm$}
\end{table}

\subsection{Objects in $\cal M$}
%
%
%
%
\begin{table}
\begin{center}
\begin{tabular}{|l|l|l|l|}  \hline
Object &   Definition &  Name  & Comments \\ \hline

$S_{ij}$ & $\frac{\Delta}{8 \pi} \left( \left[ K_{ij} \right] - g_{ij} \left[
K^i_i \right] \right)$
  & {\tt S3(dn,dn)} & intrinsic stress \\
 & & & energy tensor \\
$\left[ n_\alpha \dot{u}^\alpha \right] = - u^i u^j \left[ K_{ij} \right]$
  & & {\tt nuJumpeqn} & {\bf identity}\\
$\overline{n_\alpha \dot{u}^\alpha} =  - u^i u^j \overline{K_{ij}}$
  & & {\tt nuMeaneqn} & {\bf identity}\\
divergence (T) & $\Delta \nabla_j S^j_i =
                  \left[ T_{\alpha \beta} \frac{\partial x^\alpha}{\partial
\xi^i}
                          n^\beta \right]$ & {\tt divSTeqn(dn)} & {\bf
identity} \\

divergence (G) & $\Delta \nabla_j S^j_i = \frac{1}{8 \pi}
                  \left[ G_{\alpha \beta} \frac{\partial x^\alpha}{\partial
\xi^i}
                          n^\beta \right]$ & {\tt divSGeqn(dn)} & {\bf
identity}\\

mean & $ -S^{ij} \overline{K_{ij}} = \Delta \left[ T_{\alpha \beta} n^\alpha
n^\beta \right]$ &
           {\tt SKTeqn} & {\bf identity} \\

mean & $ -S^{ij} \overline{K_{ij}} = \frac{\Delta}{8 \pi} \left[ G_{\alpha
\beta} n^\alpha n^\beta \right]$ &
           {\tt SKGeqn} &  {\bf identity} \\ \hline
\end{tabular}
\end{center}
\caption{Objects in $\cal M$ on $\Sigma$ (General Case)}
\end{table}

\begin{table}
\begin{center}
\begin{tabular}{|l|l|l|l|}  \hline
Object &   Definition &  Name  & Comments \\ \hline

$\sigma$ & $- \Delta S_{ij} u^i u^j $ & {\tt sigma} & surface \\
 & & & energy density \\

$P$ & $(\sigma + S^i_i)/2$ & {\tt P} & isotropic \\
 & & & surface pressure \\

$\sigma_1$ & $\sigma(\xi^i)$ & {\tt sigma1} & arbitrary energy \\
 & & & density on $\Sigma$ \\

$P_1$ & $P(\xi^i)$ & {\tt P1} & arbitrary isotropic \\
 & & & pressure on $\Sigma$ \\

 & $\left[n_\alpha \dot{u}^\alpha \right] = 8 \pi (P + \sigma / 2)$
  & {\tt nuPeqn} & \\

 & $\left[n_\alpha \dot{u}^\alpha \right] = 8 \pi (P_1 + \sigma_1 / 2)$
  & {\tt nuP1eqn} & \\

conservation & $\dot{\sigma} + (\sigma + P)\Phi =
                 -\Delta \left[ T_{\alpha \beta} u^\alpha n^\beta \right]$ &
               {\tt CTeqn} &  \\
law (T) & & & \\

conservation & $\dot{\sigma} + (\sigma + P)\Phi =
                 -\Delta \left[ G_{\alpha \beta} u^\alpha n^\beta \right]/8
\pi$ &
               {\tt CGeqn} & \\
law (G) & & & \\

conservation & $\dot{\sigma_1} + (\sigma_1 + P_1)\Phi =
                 -\Delta \left[ T_{\alpha \beta} u^\alpha n^\beta \right]$ &
               {\tt C1Teqn} &  \\
law 1 (T) & & & \\

conservation & $\dot{\sigma_1} + (\sigma_1 + P_1)\Phi =
                 -\Delta \left[ G_{\alpha \beta} u^\alpha n^\beta \right]/8
\pi$ &
               {\tt C1Geqn} &  \\
law 1 (G) & & & \\

history (T) & $\Delta(\sigma + P)\overline{n_\alpha \dot{u}^\alpha} +
              P \overline{K^i_i} =
               -\Delta \left[ T_{\alpha \beta} n^\alpha n^\beta \right]$ &
              {\tt HTeqn} & \\

history (G) & $\Delta(\sigma + P)\overline{n_\alpha \dot{u}^\alpha} +
          P \overline{K^i_i} =
               -\Delta \left[ G_{\alpha \beta} n^\alpha n^\beta \right]/8 \pi$
&
              {\tt HGeqn} & \\

history (T) & $\Delta(\sigma_1 + P_1)\overline{n_\alpha \dot{u}^\alpha} +
          P_1 \overline{K^i_i} =
               -\Delta \left[ T_{\alpha \beta} n^\alpha n^\beta \right]$ &
              {\tt H1Teqn} & \\

history (G) & $\Delta(\sigma_1 + P_1)\overline{n_\alpha \dot{u}^\alpha} +
          P_1 \overline{K^i_i} =
               -\Delta \left[ G_{\alpha \beta} n^\alpha n^\beta \right]/8 \pi$
&
              {\tt H1Geqn} &       \\
evolution &
$\dot{R}^2 = \Delta + \left( \frac{\left[ m \right]}{4 \pi R^2 \sigma_1}
\right)^2
   - \frac{2 \Delta \overline{m}}{R} +
   \left( \frac{4 \pi R^2 \sigma_1}{2R} \right)^2$ &
   {\tt evInt1} & first integral of evolution \\
integral & & & equation (spherical \\
 & & & one parameter shells only) \\
evolution &
$\dot{R}^2 = \Delta + \left( \frac{\left[ m \right]}{4 \pi R^2 \sigma}
\right)^2
   - \frac{2 \Delta \overline{m}}{R} +
   \left( \frac{4 \pi R^2 \sigma}{2R} \right)^2$ &
   {\tt evInt} & first integral of evolution \\
integral & & & equation (spherical \\
 & & & one parameter shells only) \\ \hline
\end{tabular}
\end{center}
\caption{Objects in $\cal M$ on $\Sigma$ (one parameter surfaces)}
\end{table}

Once the user has loaded two 4-manifolds and specified two 3-surfaces
the command {\tt join} is used to identify the two surfaces. A
variety of objects relating to the boundary surface or shell can now be
calculated (Tables 3 and 4). \\

The user now has four active metrics in the {\tt GRTensorII} session
($\Sigma^\pm, M^\pm$). After using {\tt join} the default metric
is $\Sigma^+$. The operators {\tt Jump} and {\tt Mean} will
by default make reference to objects in $\Sigma^\pm$. In general
these operators can be used to take the jump or mean of objects
from the default metric to any user specified metric. For example
if the user had changed the default metric to $M^+$ (say
{\tt Schw}) and wished to evaluate (\ref{eqn-Lich1}) with
$M^-$ as e.g. {\tt SchwInterior} this would be done by using the
metric name of $M^-$ as a second parameter to {\tt Jump} i.e.
\verb|Jump[g(dn,dn),SchwInterior]|. \\

In practice users may wish to structure calculations and determine
results in a variety of ways and this can require minor differences
in the definitions of objects. Consider a user who wishes to study
the dynamics of a Minkowski - shell - Schwarzschild scenario.
After specifying the
two surfaces and joining them they might elect to examine the
density and pressures of the shell as given by
(\ref{eqn-sigma}) and (\ref{eqn-pressure}) which is done by referring to
the objects {\tt sigma} and {\tt P}. When considering the
dynamics they may wish to have $\sigma$ and $P$ in explicit form
or they might opt to have them appear simply as $\sigma(\xi^a)$
and $P(\xi^a)$ since this allows them to set $P(\xi^a) = 0$ to
study the dust case. Consequently there are two versions of the
history and conservation equations in Table 4
(e.g. {\tt HGeqn} vs {\tt H1Geqn}). \\

Further variations are provided to allow users to specify a stress-energy
of the 4-manifold for use in the conservation and history equations.
A user might opt to specify e.g. $T^\theta_\theta = p(r)$ instead of
using $G^\theta_\theta / 8\pi$ which might be a less transparent combination
of functions of the 4-metric. This requires the definition of
separate objects for each case (e.g. {\tt HGeqn} vs {\tt HTeqn}). \\

The implementation of (\ref{eqn-lakeid}) is complicated by the
fact that the definition requires coordinates $\theta$ and $\phi$ in
$M^\pm$ and makes reference to the shell function $R(\tau)$.
Users must adhere to these conventions (by using coordinate names
{\tt theta}, {\tt phi} and $r$ and specifying $r=R(\tau)$)
if they
wish to make use of this relation for the first integral of
evolution equation. The junction package checks for compliance
with these conventions before it will evaluate (\ref{eqn-lakeid})
to ensure that spurious results will not be generated. \\

\section{Examples}

\subsection{Joining Schwarzschild to Uniform Dust}
We begin the demonstration of the {\tt GRjunction} package by
{\em deriving~}~the junction of the Schwarzschild metric to an FRW
spacetime \cite{OppSnyder}.
(Many standard treatments merely verify that
given certain values for $m$ and $R(\tau)$ in Schwarzschild
a boundary surface exists, see e.g. \cite{MTW}). The example also illustrates
the interactive process which is required to determine
junction conditions. The pedagogic comments make the example appear
somewhat lengthy, however the example is computationally trivial
(less than 3 CPU seconds on a SUN SPARC 5). \\

We seek to join the Schwarzschild metric in the form
\begin{equation}
\label{eqn-schw}
ds^2 = \frac{dr^2}{1-2m/r} + r^2 d\theta_2^2 + r^2 \sin{\theta_2}^2 d\phi_2^2
 - (1-2m/r) {dt_2}^2
\end{equation}
to a portion of the closed FRW metric written as
\begin{equation}
\label{eqn-frw}
ds^2 = a(t)^2 (d\chi^2 + \sin{\chi}^2 (d\theta_1^2 + \sin{\theta_1}^2
d\phi_1^2))
- dt^2
\end{equation}
We take $(\theta, \phi, \tau)$ as coordinates on $\Sigma$.
The definition of the surface in (\ref{eqn-schw}) is
\begin{equation}
r = R(\tau), \theta_2 = \theta, \phi_2 = \phi, t_2 = T(\tau)
\end{equation}
with $r-R(\tau) = 0$ as the parametric form for the surface. For the
FRW metric the surface is specified as
\begin{equation}
\chi = X, \theta_1 = \theta, \phi_1 = \phi, t = \tau
\end{equation}
with parametric form $\chi - X = 0$. \\

The following session output (from this point to the end
of this subsection) was taken directly from the Latex output
provided by Maple. \\

\noindent We begin by loading the GRTensorII library into Maple.

\begin{mapleinput}
readlib(grii):
\end{mapleinput}
\noindent Next we load the junction package into GRTensorII.

\begin{mapleinput}
grlib(junction);
\end{mapleinput}
\begin{maplettyout}
The GRJunction Package
Last modified September 28, 1995
Developed by Peter Musgrave and Kayll Lake, (c) 1995

\end{maplettyout}
\noindent The first step is to load the Schwarzschild metric via {\tt qload}.

\begin{mapleinput}
qload(Schw);
\end{mapleinput}
\begin{maplelatex}
\[
{\it Default\:spacetime}={\it Schw}
\]
\end{maplelatex}
\begin{maplelatex}
\[
{\it For\:the\:Schw\:spacetime:}
\]
\end{maplelatex}
\begin{maplelatex}
\[
{\it Coordinates}
\]
\end{maplelatex}
\begin{maplelatex}
\[
{\it x\:}^{{1}}={r}, {\it x\:}^{{2}}={{ \theta}_{2}}, {\it x\:}^{
{3}}={{ \phi}_{2}}, {\it x\:}^{{4}}={{t}_{2}}
\]
\end{maplelatex}
\begin{maplelatex}
\[
{\it Line\:element}
\]
\end{maplelatex}
\begin{maplelatex}
\[
{\it \:ds}^{2}={\displaystyle \frac {{\it \:d}\,{r}^{{\it 2\:}}}{
1 - 2\,{\displaystyle \frac {{m}}{{r}}}}} + {r}^{2}\,{\it \:d}\,{
{ \theta}_{2}}^{{\it 2\:}} + {r}^{2}\,{\rm sin} \left( \! \,{{
\theta}_{2}}\, \!  \right) ^{2}\,{\it \:d}\,{{ \phi}_{2}}^{{\it 2
\:}} +  \left( \! \, - 1 + 2\,{\displaystyle \frac {{m}}{{r}}}\,
 \!  \right) \,{\it \:d}\,{{t}_{2}}^{{\it 2\:}}
\]
\end{maplelatex}
\noindent Now we use the {\tt surf} command to specify a surface. {\tt surf}
prompts the user for information
which defines the surface and its normal vector.

\begin{mapleinput}
surf(Schw,ssurf);
\end{mapleinput}
\noindent First we are asked for the coordinates on the surface as a list.

\begin{maplettyout}
Please enter the coordinates of the surface as a list
e.g. [theta, phi, tau];
Enter a list >
\end{maplettyout}
\begin{mapleinput}
[theta,phi,tau];
\end{mapleinput}
\noindent Next we are asked to specify which of the $\xi^i$ is the parameter
for a one-parameter shell
(if any).

\begin{maplettyout}
For a one-parameter shell enter the parameter (0 for none) >
\end{maplettyout}
\begin{mapleinput}
tau;
\end{mapleinput}
\noindent Now we provide the coordinate definition of $\Sigma$.
(Note that in Maple $\left[ ~~ \right]$'s do double duty; they
denote lists and are used to indicate subscripts)

\begin{maplettyout}

Please enter the coordinate definition of the surface
(the x{^a} = x(xi{^b})  ) as a LIST.
e.g. [ r=R(tau), theta=theta, phi=phi, t=T(tau)]
 >
\end{maplettyout}
\begin{mapleinput}
[r=R(tau), theta[2]=theta, phi[2]=phi, t[2]=T(tau) ];
\end{mapleinput}
\begin{maplelatex}
\[
{\it CPU\:Time\:}=.050
\]
\end{maplelatex}
\noindent The character of the normal vector is now entered.

\begin{maplettyout}
Please indicate the nature of the surface normal vector
(-1 = timelike, 1= spacelike) Enter +1,0 or -1 >
\end{maplettyout}
\begin{mapleinput}
1;
\end{mapleinput}
\noindent Since we have provided a shell parameter  we are given the option of
employing the relation
$u_\alpha u^\alpha = -1$ to eliminate the derivative of one of the quantities
used
in the coordinate definition of the surface. We choose to eliminate
$\partial T(\tau)/ \partial \tau$ (in Maple parlance {\tt diff(T(tau),tau)}).
This
choice will produce $d\tau^2 = -1$ on $\Sigma$.

\begin{maplettyout}
Use  +/- 1 = u{^a} u{a} as a constraint ? Enter 1 if yes >
\end{maplettyout}
\begin{mapleinput}
1;
\end{mapleinput}
\begin{maplettyout}
Created definition for u(dn)

\end{maplettyout}
\begin{maplelatex}
\[
{\it CPU\:Time\:}=.100
\]
\end{maplelatex}
\begin{maplelatex}
\[
{\it The\:constraint\:equation\:is:}
\]
\end{maplelatex}
\begin{maplelatex}
\[
-1={\displaystyle \frac { - {r}^{2}\, \left( \! \,{\frac {{
\partial}}{{ \partial}{ \tau}}}\,{\rm R}(\,{ \tau}\,)\, \!
 \right) ^{2} +  \left( \! \,{\frac {{ \partial}}{{ \partial}{
\tau}}}\,{\rm T}(\,{ \tau}\,)\, \!  \right) ^{2}\,{r}^{2} - 4\,
 \left( \! \,{\frac {{ \partial}}{{ \partial}{ \tau}}}\,{\rm T}(
\,{ \tau}\,)\, \!  \right) ^{2}\,{r}\,{m} + 4\, \left( \! \,
{\frac {{ \partial}}{{ \partial}{ \tau}}}\,{\rm T}(\,{ \tau}\,)\,
 \!  \right) ^{2}\,{m}^{2}}{(\, - {r} + 2\,{m}\,)\,{r}}}
\]
\end{maplelatex}
\begin{maplettyout}
Enter the term you wish to use the constraint to eliminate
Term >
\end{maplettyout}
\begin{mapleinput}
diff(T(tau),tau);
\end{mapleinput}
\noindent In solving for $\partial T(\tau)/ \partial \tau$ Maple determines
that there are two choices
(differing in sign). We are asked to choose one of them.

\begin{maplettyout}
Solve returned multiple solutions. They are:

\end{maplettyout}
\begin{maplelatex}
\[
{\it [1]}, {\displaystyle \frac {\sqrt {{\displaystyle \frac {{r}
 - 2\,{m} + {r}\, \left( \! \,{\frac {{ \partial}}{{ \partial}{
\tau}}}\,{\rm R}(\,{ \tau}\,)\, \!  \right) ^{2}}{{r}}}}}{ - \,
{\displaystyle \frac {{r}}{ - {r} + 2\,{m}}} + 4\,{\displaystyle
\frac {{m}}{ - {r} + 2\,{m}}} - 4\,{\displaystyle \frac {{m}^{2}
}{(\, - {r} + 2\,{m}\,)\,{r}}}}}
\]
\end{maplelatex}
\begin{maplelatex}
\[
{\it [2]},  - \,{\displaystyle \frac {\sqrt {{\displaystyle
\frac {{r} - 2\,{m} + {r}\, \left( \! \,{\frac {{ \partial}}{{
\partial}{ \tau}}}\,{\rm R}(\,{ \tau}\,)\, \!  \right) ^{2}}{{r}
}}}}{ - \,{\displaystyle \frac {{r}}{ - {r} + 2\,{m}}} + 4\,
{\displaystyle \frac {{m}}{ - {r} + 2\,{m}}} - 4\,{\displaystyle
\frac {{m}^{2}}{(\, - {r} + 2\,{m}\,)\,{r}}}}}
\]
\end{maplelatex}
\begin{maplettyout}

Please select a solution.
Enter choice >
\end{maplettyout}
\begin{mapleinput}
1;
\end{mapleinput}
\noindent Now we enter the parametric definition of the surface.

\begin{maplettyout}
Default metric is now Schw
Please enter the equation for the surface.
The surface will be defined by setting the
equation you enter to zero.

\end{maplettyout}
\begin{maplelatex}
\[
{\it surface}
\]
\end{maplelatex}
\begin{maplettyout}
 >
\end{maplettyout}
\begin{mapleinput}
r-R(t[2]);
\end{mapleinput}
\begin{maplelatex}
\[
{\it CPU\:Time\:}=.083
\]
\end{maplelatex}
\noindent Finally, we are asked explicitly for the sign of the normal vector.

\begin{maplettyout}
The definition of the normal vector is +/- grad(surface)
please enter +1 or -1 to indicate the CHOICE of sign
Enter +1,-1 >
\end{maplettyout}
\begin{mapleinput}
1;
\end{mapleinput}
\begin{maplelatex}
\[
{\it CPU\:Time\:}=.184
\]
\end{maplelatex}
\begin{maplettyout}
Default metric is now ssurf

\end{maplettyout}
\begin{maplelatex}
\[
{\it CPU\:Time\:}=.133
\]
\end{maplelatex}
\begin{maplelatex}
\[
{\it For\:the\:Schw\:spacetime:}
\]
\end{maplelatex}
\begin{maplelatex}
\[
{\it The\:Equation\:of\:the\:surface}
\]
\end{maplelatex}
\begin{maplelatex}
\[
{\it surface\:}={r} - {\rm R} \left( \! \,{{t}_{2}}\, \!
 \right)
\]
\end{maplelatex}
\begin{maplelatex}
\[
{\it For\:the\:Schw\:spacetime:}
\]
\end{maplelatex}
\begin{maplelatex}
\[
{\it Coordinate\:transforms\:onto\:the\:surface}
\]
\end{maplelatex}
\begin{maplelatex}
\[
{\it xform\:}^{{r}}={\rm R}(\,{ \tau}\,)
\]
\end{maplelatex}
\begin{maplelatex}
\[
{\it xform\:}^{{{ \theta}_{2}}}={ \theta}
\]
\end{maplelatex}
\begin{maplelatex}
\[
{\it xform\:}^{{{ \phi}_{2}}}={ \phi}
\]
\end{maplelatex}
\begin{maplelatex}
\[
{\it xform\:}^{{{t}_{2}}}={\rm T}(\,{ \tau}\,)
\]
\end{maplelatex}
\begin{maplelatex}
\[
{\it For\:the\:ssurf\:spacetime:}
\]
\end{maplelatex}
\begin{maplelatex}
\[
{\it Line\:element}
\]
\end{maplelatex}
\begin{maplelatex}
\[
{\it \:ds}^{2}={\rm R}(\,{ \tau}\,)^{2}\,{\it \:d}\,{ \theta}^{
{\it 2\:}} + {\rm R}(\,{ \tau}\,)^{2}\,{\rm sin}(\,{ \theta}\,)^{
2}\,{\it \:d}\,{ \phi}^{{\it 2\:}} - {\it \:d}\,{ \tau}^{{\it 2\:
}}
\]
\end{maplelatex}
\begin{maplettyout}
The intrinsic metric and normal vector have been calculated.
You may wish to simplify them further before saving the surface
or calculating K(dn,dn)

\end{maplettyout}
\noindent This completes the specification of the surface in the Schwarzschild
exterior.
Next we  load the FRW metric (which we will take as the interior).

\begin{mapleinput}
qload(frw);
\end{mapleinput}
\begin{maplelatex}
\[
{\it Default\:spacetime}={\it frw}
\]
\end{maplelatex}
\begin{maplelatex}
\[
{\it For\:the\:frw\:spacetime:}
\]
\end{maplelatex}
\begin{maplelatex}
\[
{\it Coordinates}
\]
\end{maplelatex}
\begin{maplelatex}
\[
{\it x\:}^{{1}}={ \chi}, {\it x\:}^{{2}}={{ \theta}_{1}}, {\it x
\:}^{{3}}={{ \phi}_{1}}, {\it x\:}^{{4}}={t}
\]
\end{maplelatex}
\begin{maplelatex}
\[
{\it Line\:element}
\]
\end{maplelatex}
\begin{maplelatex}
\[
{\it \:ds}^{2}={\rm a}(\,{t}\,)^{2}\,{\it \:d}\,{ \chi}^{{\it 2\:
}} + {\rm a}(\,{t}\,)^{2}\,{\rm sin}(\,{ \chi}\,)^{2}\,{\it \:d}
\,{{ \theta}_{1}}^{{\it 2\:}} + {\rm a}(\,{t}\,)^{2}\,{\rm sin}(
\,{ \chi}\,)^{2}\,{\rm sin} \left( \! \,{{ \theta}_{1}}\, \!
 \right) ^{2}\,{\it \:d}\,{{ \phi}_{1}}^{{\it 2\:}} - {\it \:d}\,
{t}^{{\it 2\:}}
\]
\end{maplelatex}
\noindent Before defining the surface in the FRW spacetime we first calculate
$G_\alpha^\beta$ for later use.

\begin{mapleinput}
grcalc(G(dn,up));
\end{mapleinput}
\begin{maplettyout}
Created definition for G(dn,up)

\end{maplettyout}
\begin{maplelatex}
\[
{\it CPU\:Time\:}=.634
\]
\end{maplelatex}
\noindent We specify the surface in $M^-$ in parametric from as $\chi - X = 0$.
The process is identical
to that followed for the Schwarzschild case above (except we do not use
$u_\alpha u^\alpha = -1$).

\begin{mapleinput}
surf(frw, fsurf);
\end{mapleinput}
\begin{maplettyout}
Please enter the coordinates of the surface as a list
e.g. [theta, phi, tau];
Enter a list >
\end{maplettyout}
\begin{mapleinput}
[theta,phi,tau];
\end{mapleinput}
\begin{maplettyout}
For a one-parameter shell enter the parameter (0 for none) >
\end{maplettyout}
\begin{mapleinput}
tau;
\end{mapleinput}
\begin{maplettyout}

Please enter the coordinate definition of the surface
(the x{^a} = x(xi{^b})  ) as a LIST.
e.g. [ r=R(tau), theta=theta, phi=phi, t=T(tau)]
 >
\end{maplettyout}
\begin{mapleinput}
[ chi = X, theta[1]=theta, phi[1]=phi, t=tau];
\end{mapleinput}
\begin{maplelatex}
\[
{\it CPU\:Time\:}=.050
\]
\end{maplelatex}
\begin{maplettyout}
Please indicate the nature of the surface normal vector
(-1 = timelike, 1= spacelike) Enter +1,0 or -1 >
\end{maplettyout}
\begin{mapleinput}
1;
\end{mapleinput}
\begin{maplettyout}
Use  +/- 1 = u{^a} u{a} as a constraint ? Enter 1 if yes >
\end{maplettyout}
\begin{mapleinput}
0;
\end{mapleinput}
\begin{maplettyout}
Default metric is now frw
Please enter the equation for the surface.
The surface will be defined by setting the
equation you enter to zero.

\end{maplettyout}
\begin{maplelatex}
\[
{\it surface}
\]
\end{maplelatex}
\begin{maplettyout}
 >
\end{maplettyout}
\begin{mapleinput}
chi-X;
\end{mapleinput}
\begin{maplelatex}
\[
{\it CPU\:Time\:}=.016
\]
\end{maplelatex}
\begin{maplettyout}
The definition of the normal vector is +/- grad(surface)
please enter +1 or -1 to indicate the CHOICE of sign
Enter +1,-1 >
\end{maplettyout}
\begin{mapleinput}
1;
\end{mapleinput}
\begin{maplelatex}
\[
{\it CPU\:Time\:}=.050
\]
\end{maplelatex}
\begin{maplettyout}
Default metric is now fsurf

\end{maplettyout}
\begin{maplelatex}
\[
{\it CPU\:Time\:}=.066
\]
\end{maplelatex}
\begin{maplelatex}
\[
{\it For\:the\:frw\:spacetime:}
\]
\end{maplelatex}
\begin{maplelatex}
\[
{\it The\:Equation\:of\:the\:surface}
\]
\end{maplelatex}
\begin{maplelatex}
\[
{\it surface\:}={ \chi} - {X}
\]
\end{maplelatex}
\begin{maplelatex}
\[
{\it For\:the\:frw\:spacetime:}
\]
\end{maplelatex}
\begin{maplelatex}
\[
{\it Coordinate\:transforms\:onto\:the\:surface}
\]
\end{maplelatex}
\begin{maplelatex}
\[
{\it xform\:}^{{ \chi}}={X}
\]
\end{maplelatex}
\begin{maplelatex}
\[
{\it xform\:}^{{{ \theta}_{1}}}={ \theta}
\]
\end{maplelatex}
\begin{maplelatex}
\[
{\it xform\:}^{{{ \phi}_{1}}}={ \phi}
\]
\end{maplelatex}
\begin{maplelatex}
\[
{\it xform\:}^{{t}}={ \tau}
\]
\end{maplelatex}
\begin{maplelatex}
\[
{\it For\:the\:fsurf\:spacetime:}
\]
\end{maplelatex}
\begin{maplelatex}
\[
{\it Line\:element}
\]
\end{maplelatex}
\begin{maplelatex}
\[
{\it \:ds}^{2}={\rm a}(\,{ \tau}\,)^{2}\,{\rm sin}(\,{X}\,)^{2}\,
{\it \:d}\,{ \theta}^{{\it 2\:}} + {\rm a}(\,{ \tau}\,)^{2}\,
{\rm sin}(\,{X}\,)^{2}\,{\rm sin}(\,{ \theta}\,)^{2}\,{\it \:d}\,
{ \phi}^{{\it 2\:}} - {\it \:d}\,{ \tau}^{{\it 2\:}}
\]
\end{maplelatex}
\begin{maplettyout}
The intrinsic metric and normal vector have been calculated.
You may wish to simplify them further before saving the surface
or calculating K(dn,dn)

\end{maplettyout}
\noindent Now we can identify the two surfaces we have specified by means of
the command {\tt join}.
By convention the first surface name in the {\tt join} command  is taken as
$\Sigma^+$ for the purposes of
evaluating e.g. $\left[ g_{ij} \right]$.

\begin{mapleinput}
join(ssurf,fsurf);
\end{mapleinput}
\begin{maplettyout}
ssurf and fsurf are now joined.
The default metric name is ssurf.
The exterior metric is: ssurf
The interior metric is: fsurf

\end{maplettyout}
\begin{maplelatex}
\[
{\it CPU\:Time\:}=.034
\]
\end{maplelatex}
\begin{maplelatex}
\[
{\it For\:the\:ssurf\:spacetime:}
\]
\end{maplelatex}
\begin{maplelatex}
\[
{\it \:Jump\:from\:defaultMetric\:-\:Mint}
\]
\end{maplelatex}
\begin{maplelatex}
\[
{{\it Jump\:[g(dn,dn),fsurf]}_{{ \theta}}}\,{{}_{{ \theta}}}=
{\rm R}(\,{ \tau}\,)^{2} - {\rm a}(\,{ \tau}\,)^{2}\,{\rm sin}(\,
{X}\,)^{2}
\]
\end{maplelatex}
\begin{maplelatex}
\[
{{\it Jump\:[g(dn,dn),fsurf]}_{{ \phi}}}\,{{}_{{ \phi}}}={\rm R}(
\,{ \tau}\,)^{2}\,{\rm sin}(\,{ \theta}\,)^{2} - {\rm a}(\,{ \tau
}\,)^{2}\,{\rm sin}(\,{X}\,)^{2}\,{\rm sin}(\,{ \theta}\,)^{2}
\]
\end{maplelatex}
\noindent To obtain $\left[ g_{ij} \right] = 0$ we require a particular value
for $R(\tau)$. We use
the GRTensor command {\tt grcomponent} to extract $\left[ g_{11} \right]$
(which we assign to
{\tt jump\_g11}). We then set {\tt jump\_g11} equal to zero and solve for
$R(\tau)$.

\begin{mapleinput}
jump_g11 := grcomponent(Jump[g(dn,dn)], [1,1]);
\end{mapleinput}
\begin{maplelatex}
\[
{\it jump\_g11} := {\rm R}(\,{ \tau}\,)^{2} - {\rm a}(\,{ \tau}\,
)^{2}\,{\rm sin}(\,{X}\,)^{2}
\]
\end{maplelatex}
\begin{mapleinput}
sol := [solve(jump_g11 = 0,R(tau))];
\end{mapleinput}
\begin{maplelatex}
\[
{\it sol} := [\,{\rm a}(\,{ \tau}\,)\,{\rm sin}(\,{X}\,),  -
{\rm a}(\,{ \tau}\,)\,{\rm sin}(\,{X}\,)\,]
\]
\end{maplelatex}
\noindent We now substitute the positive solution back in to verify $\left[
g_{ij} \right] = 0$. This is accomplished
by using the routine {\tt grmap} to map the Maple substitution command {\tt
subs} over the
components of \verb|Jump[g(dn,dn)]|. (The \verb|'x'| is a placeholder
indicating which of the
arguments to {\tt subs} is to be filled in by the component value.)

\begin{mapleinput}
grmap(Jump[g(dn,dn)], subs, R(tau) = sol[1], 'x');
\end{mapleinput}
\begin{maplettyout}
Applying routine subs to Jump[g(dn,dn),fsurf]

\end{maplettyout}
\noindent (Here we use a GRTensor short-cut. The \_ refers to the last
mentioned object. In this
case \verb|Jump[g(dn,dn)]| allowing us to save some typing).

\begin{mapleinput}
grdisplay(_);
\end{mapleinput}
\begin{maplelatex}
\[
{\it For\:the\:ssurf\:spacetime:}
\]
\end{maplelatex}
\begin{maplelatex}
\[
{\it \:Jump\:from\:defaultMetric\:-\:Mint}
\]
\end{maplelatex}
\begin{maplelatex}
\[
{\it Jump[g(dn,dn),fsurf]}={\it All\:components\:are\:zero}
\]
\end{maplelatex}
\noindent To establish a Schwarzschild-Dust boundary surface we next need to
establish that
$\left[ K_{ij} \right] = 0$. We begin by calculating \verb|Jump[K(dn,up)]|. We
prefer
the mixed form $K_i^j$ so that we can make later use of  $G_\alpha^\beta$ and
the
phenomenology of the FRW space.

\begin{mapleinput}
grcalc(Jump[K(dn,up)]);
\end{mapleinput}
\begin{maplettyout}
Created definition for K(dn,up)

\end{maplettyout}
\begin{maplelatex}
\[
{\it CPU\:Time\:}=1.183
\]
\end{maplelatex}
\begin{mapleinput}
grdisplay(_);
\end{mapleinput}
\begin{maplelatex}
\[
{\it For\:the\:ssurf\:spacetime:}
\]
\end{maplelatex}
\begin{maplelatex}
\[
{\it \:Jump\:from\:defaultMetric\:-\:Mint}
\]
\end{maplelatex}
\begin{maplelatex}
\begin{eqnarray*}
\lefteqn{{{\it Jump\:[K(dn,up),fsurf]}_{{ \theta}}}\,{}^{{ \theta
}}=} \\
 & &  - \,{\displaystyle \frac { - \sqrt {{\displaystyle \frac {
{\rm R}(\,{ \tau}\,) - 2\,{m} + {\rm R}(\,{ \tau}\,)\, \left( \!
\,{\frac {{ \partial}}{{ \partial}{ \tau}}}\,{\rm R}(\,{ \tau}\,)
\, \!  \right) ^{2}}{{\rm R}(\,{ \tau}\,)}}}\,{\rm a}(\,{ \tau}\,
)\,{\rm sin}(\,{X}\,) + {\rm cos}(\,{X}\,)\,{\rm R}(\,{ \tau}\,)
}{{\rm R}(\,{ \tau}\,)\,{\rm a}(\,{ \tau}\,)\,{\rm sin}(\,{X}\,)
}}
\end{eqnarray*}
\end{maplelatex}
\begin{maplelatex}
\begin{eqnarray*}
\lefteqn{{{\it Jump\:[K(dn,up),fsurf]}_{{ \phi}}}\,{}^{{ \phi}}=}
 \\
 & &  - \,{\displaystyle \frac { - \sqrt {{\displaystyle \frac {
{\rm R}(\,{ \tau}\,) - 2\,{m} + {\rm R}(\,{ \tau}\,)\, \left( \!
\,{\frac {{ \partial}}{{ \partial}{ \tau}}}\,{\rm R}(\,{ \tau}\,)
\, \!  \right) ^{2}}{{\rm R}(\,{ \tau}\,)}}}\,{\rm a}(\,{ \tau}\,
)\,{\rm sin}(\,{X}\,) + {\rm cos}(\,{X}\,)\,{\rm R}(\,{ \tau}\,)
}{{\rm R}(\,{ \tau}\,)\,{\rm a}(\,{ \tau}\,)\,{\rm sin}(\,{X}\,)
}}
\end{eqnarray*}
\end{maplelatex}
\begin{maplelatex}
\[
{{\it Jump\:[K(dn,up),fsurf]}_{{ \tau}}}\,{}^{{ \tau}}=
{\displaystyle \frac {{\rm R}(\,{ \tau}\,)^{2}\, \left( \! \,
{\frac {{ \partial}^{2}}{{ \partial}{ \tau}^{2}}}\,{\rm R}(\,{
\tau}\,)\, \!  \right)  + {m}}{\sqrt {{\displaystyle \frac {{\rm
R}(\,{ \tau}\,) - 2\,{m} + {\rm R}(\,{ \tau}\,)\, \left( \! \,
{\frac {{ \partial}}{{ \partial}{ \tau}}}\,{\rm R}(\,{ \tau}\,)\,
 \!  \right) ^{2}}{{\rm R}(\,{ \tau}\,)}}}\,{\rm R}(\,{ \tau}\,)
^{2}}}
\]
\end{maplelatex}
\noindent Setting   $\left[ K_\tau^\tau \right] = 0$ gives a value for the
Schwarzschild mass $m$ in terms
of $R(\tau)$ and we substitute in this value. This leaves only the angular jump
in $K$ to contend
with.

\begin{mapleinput}
grmap(_, subs, m = -R(tau)^2*diff(R(tau),tau,tau), 'x');
\end{mapleinput}
\begin{maplettyout}
Applying routine subs to Jump[K(dn,up),fsurf]

\end{maplettyout}
\noindent Once again we make use of the value of $R(\tau)$ which was required
for $\left[ g_{ij} \right] = 0$.

\begin{mapleinput}
grmap(_,subs, R(tau) = sol[1],  'x');
\end{mapleinput}
\begin{maplettyout}
Applying routine subs to Jump[K(dn,up),fsurf]

\end{maplettyout}
\begin{mapleinput}
gralter(_,factor);
\end{mapleinput}
\begin{maplettyout}
Component Alteration of a grtensor object:

Applying routine factor to object Jump[K(dn,up),fsurf]

\end{maplettyout}
\begin{maplelatex}
\[
{\it CPU\:Time\:}=.150
\]
\end{maplelatex}
\begin{mapleinput}
grdisplay(_);
\end{mapleinput}
\begin{maplelatex}
\[
{\it For\:the\:ssurf\:spacetime:}
\]
\end{maplelatex}
\begin{maplelatex}
\[
{\it \:Jump\:from\:defaultMetric\:-\:Mint}
\]
\end{maplelatex}
\begin{maplelatex}
\begin{eqnarray*}
\lefteqn{{{\it Jump\:[K(dn,up),fsurf]}_{{ \theta}}}\,{}^{{ \theta
}}=} \\
 & &  - \,{\displaystyle \frac { - \sqrt {1 + 2\,{\rm a}(\,{ \tau
}\,)\,{\rm sin}(\,{X}\,)^{2}\, \left( \! \,{\frac {{ \partial}^{2
}}{{ \partial}{ \tau}^{2}}}\,{\rm a}(\,{ \tau}\,)\, \!  \right)
 + {\rm sin}(\,{X}\,)^{2}\, \left( \! \,{\frac {{ \partial}}{{
\partial}{ \tau}}}\,{\rm a}(\,{ \tau}\,)\, \!  \right) ^{2}} +
{\rm cos}(\,{X}\,)}{{\rm a}(\,{ \tau}\,)\,{\rm sin}(\,{X}\,)}}
\end{eqnarray*}
\end{maplelatex}
\begin{maplelatex}
\begin{eqnarray*}
\lefteqn{{{\it Jump\:[K(dn,up),fsurf]}_{{ \phi}}}\,{}^{{ \phi}}=}
 \\
 & &  - \,{\displaystyle \frac { - \sqrt {1 + 2\,{\rm a}(\,{ \tau
}\,)\,{\rm sin}(\,{X}\,)^{2}\, \left( \! \,{\frac {{ \partial}^{2
}}{{ \partial}{ \tau}^{2}}}\,{\rm a}(\,{ \tau}\,)\, \!  \right)
 + {\rm sin}(\,{X}\,)^{2}\, \left( \! \,{\frac {{ \partial}}{{
\partial}{ \tau}}}\,{\rm a}(\,{ \tau}\,)\, \!  \right) ^{2}} +
{\rm cos}(\,{X}\,)}{{\rm a}(\,{ \tau}\,)\,{\rm sin}(\,{X}\,)}}
\end{eqnarray*}
\end{maplelatex}
\noindent At this point we need to make reference to the Einstein tensor for
the FRW spacetime. To this
point we have not imposed the restriction that the interior be dust. We do this
by setting
$G_\theta^\theta = 0$ and solving for $\partial^2 a(t) / \partial t^2$. This is
then used
in \verb|Jump[K(dn,up)]| (with $t = \tau$ on the surface as required). Recall
that by default object
names refer to the {\tt ssurf} spacetime (i.e $\Sigma^+$). The use of a metric
name in
square brackets after the tensor name below indicates which metric the object
is to be taken
from.

\begin{mapleinput}
grdisplay(G[frw](dn,up));
\end{mapleinput}
\begin{maplelatex}
\[
{\it For\:the\:frw\:spacetime:}
\]
\end{maplelatex}
\begin{maplelatex}
\[
{\it G(dn,up)}
\]
\end{maplelatex}
\begin{maplelatex}
\[
{{\it G\:}_{{ \chi}}}\,{}^{{ \chi}}= - \,{\displaystyle \frac {
{\rm sin}(\,{ \chi}\,)^{2}\, \left( \! \,{\frac {{ \partial}}{{
\partial}{t}}}\,{\rm a}(\,{t}\,)\, \!  \right) ^{2} + 2\,{\rm a}(\,{t}\,)\,{\rm
sin}(\,{ \chi}\,)^{2}\, \left( \! \,{\frac {{
\partial}^{2}}{{ \partial}{t}^{2}}}\,{\rm a}(\,{t}\,)\, \!
 \right)  + 1 - {\rm cos}(\,{ \chi}\,)^{2}}{{\rm a}(\,{t}\,)^{2}\,{\rm sin}(\,{
\chi}\,)^{2}}}
\]
\end{maplelatex}
\begin{maplelatex}
\[
{{\it G\:}_{{{ \theta}_{1}}}}\,{}^{{{ \theta}_{1}}}= - \,
{\displaystyle \frac {1 +  \left( \! \,{\frac {{ \partial}}{{
\partial}{t}}}\,{\rm a}(\,{t}\,)\, \!  \right) ^{2} + 2\,{\rm a}(
\,{t}\,)\, \left( \! \,{\frac {{ \partial}^{2}}{{ \partial}{t}^{2
}}}\,{\rm a}(\,{t}\,)\, \!  \right) }{{\rm a}(\,{t}\,)^{2}}}
\]
\end{maplelatex}
\begin{maplelatex}
\[
{{\it G\:}_{{{ \phi}_{1}}}}\,{}^{{{ \phi}_{1}}}= - \,
{\displaystyle \frac {1 +  \left( \! \,{\frac {{ \partial}}{{
\partial}{t}}}\,{\rm a}(\,{t}\,)\, \!  \right) ^{2} + 2\,{\rm a}(
\,{t}\,)\, \left( \! \,{\frac {{ \partial}^{2}}{{ \partial}{t}^{2
}}}\,{\rm a}(\,{t}\,)\, \!  \right) }{{\rm a}(\,{t}\,)^{2}}}
\]
\end{maplelatex}
\begin{maplelatex}
\[
{{\it G\:}_{{t}}}\,{}^{{t}}= - \,{\displaystyle \frac {2\,{\rm
sin}(\,{ \chi}\,)^{2} + 3\,{\rm sin}(\,{ \chi}\,)^{2}\, \left(
\! \,{\frac {{ \partial}}{{ \partial}{t}}}\,{\rm a}(\,{t}\,)\,
 \!  \right) ^{2} + 1 - {\rm cos}(\,{ \chi}\,)^{2}}{{\rm a}(\,{t}
\,)^{2}\,{\rm sin}(\,{ \chi}\,)^{2}}}
\]
\end{maplelatex}
\noindent To make use of the condition $p=0$ we will extract the
$G_{\theta_1}^{\theta_1}$
component (using {\tt grcomponent}) and set it equal to zero and then  isolate
the $\partial^2 a(t)/\partial t^2$ term. We will then substitute this into
$\left[ K_i^j \right]$.
(We need the Maple {\tt isolate} library which we we now load)

\begin{mapleinput}
readlib(isolate):
\end{mapleinput}
The $\partial^2 a(t) / \partial t^2$ term is now isolated and we change $t$ to
$\tau$ (via the {\tt subs} command).

\begin{mapleinput}
da := subs( t=tau, isolate( grcomponent(G[frw](dn,up),[2,2]) = 0,
diff(a(t),t$2) ));
\end{mapleinput}
\begin{maplelatex}
\[
{\it da} := {\frac {{ \partial}^{2}}{{ \partial}{ \tau}^{2}}}\,
{\rm a}(\,{ \tau}\,)={\displaystyle \frac {1}{2}}\,
{\displaystyle \frac { - 1 -  \left( \! \,{\frac {{ \partial}}{{
\partial}{ \tau}}}\,{\rm a}(\,{ \tau}\,)\, \!  \right) ^{2}}{
{\rm a}(\,{ \tau}\,)}}
\]
\end{maplelatex}
\noindent The resulting equation {\tt da} is substituted into $\left[ K_i^j
\right]$ and some simplification is then performed.

\begin{mapleinput}
grmap( Jump[K(dn,up)], subs, da, 'x');
\end{mapleinput}
\begin{maplettyout}
Applying routine subs to Jump[K(dn,up),fsurf]

\end{maplettyout}
\begin{mapleinput}
gralter(_,expand,trig);
\end{mapleinput}
\begin{maplettyout}
Component Alteration of a grtensor object:

Applying routine expand to object Jump[K(dn,up),fsurf]
Applying routine simplify[trig] to object Jump[K(dn,up),fsurf]

\end{maplettyout}
\begin{maplelatex}
\[
{\it CPU\:Time\:}=.267
\]
\end{maplelatex}
\begin{mapleinput}
grdisplay(_);
\end{mapleinput}
\begin{maplelatex}
\[
{\it For\:the\:ssurf\:spacetime:}
\]
\end{maplelatex}
\begin{maplelatex}
\[
{\it \:Jump\:from\:defaultMetric\:-\:Mint}
\]
\end{maplelatex}
\begin{maplelatex}
\[
{{\it Jump\:[K(dn,up),fsurf]}_{{ \theta}}}\,{}^{{ \theta}}= - \,
{\displaystyle \frac { - \sqrt {{\rm cos}(\,{X}\,)^{2}} + {\rm
cos}(\,{X}\,)}{{\rm a}(\,{ \tau}\,)\,{\rm sin}(\,{X}\,)}}
\]
\end{maplelatex}
\begin{maplelatex}
\[
{{\it Jump\:[K(dn,up),fsurf]}_{{ \phi}}}\,{}^{{ \phi}}= - \,
{\displaystyle \frac { - \sqrt {{\rm cos}(\,{X}\,)^{2}} + {\rm
cos}(\,{X}\,)}{{\rm a}(\,{ \tau}\,)\,{\rm sin}(\,{X}\,)}}
\]
\end{maplelatex}
\noindent We're nearly there, but Maple will not collapse e.g. $\sqrt{x^2}$ to
$x$ unless it is sure
$x$ is real, or explicitly told to do so. We tell it to go ahead by using the
routine
\verb|simplify[sqrt,symbolic]|.

\begin{mapleinput}
gralter(_,sqrt);
\end{mapleinput}
\begin{maplettyout}
Component Alteration of a grtensor object:

Applying routine simplify[sqrt] to object Jump[K(dn,up),fsurf]

\end{maplettyout}
\begin{maplelatex}
\[
{\it CPU\:Time\:}=.066
\]
\end{maplelatex}
\noindent So if we take the FRW solution to be dust then we can establish that
$\left[ g_{ij} \right] = 0$ and $\left[ K_i^j \right] = 0$ completing the
junction of Schwarzschild to FRW.

\begin{mapleinput}
grdisplay(_);
\end{mapleinput}
\begin{maplelatex}
\[
{\it For\:the\:ssurf\:spacetime:}
\]
\end{maplelatex}
\begin{maplelatex}
\[
{\it \:Jump\:from\:defaultMetric\:-\:Mint}
\]
\end{maplelatex}
\begin{maplelatex}
\[
{\it Jump[K(dn,up),fsurf]}={\it All\:components\:are\:zero}
\]
\end{maplelatex}
\noindent With the FRW solution restricted to uniform dust then we can
establish that
$\left[ g_{ij} \right] = 0$ and $\left[ K_i^j \right] = 0$ and hence a boundary
surface
exists provided we take $R(\tau) = a(\tau) \sin{X}$
and $m = - R(\tau)^2 \partial^2 R(\tau)/ \partial r^2$ in Schwarzschild.


\subsection{Shells in Spherically Symmetric Static Spacetimes}

The evolution of thin shells in spherically symmetric spacetimes
has been widely studied. Here we demonstrate how {\tt GRJunction}
is employed to determine the evolution of a thin-shell separating
two spherically symmetric static spacetimes. This result
contains as special case the classic analysis of Israel \cite{Israel66}. \\

We take $M^+$ and $M^-$ as
\begin{eqnarray}
\label{eqn-staticf}
ds_-^2 & = & dr_-^2/f(r_-) + r_-^2 d\Omega^2 - f(r_-) dt_1^2, \\
\label{eqn-staticF}
ds_+^2 & = & dr_+^2/F(r_+) + r_+^2 d\Omega^2 - F(r_+) dt_2^2
\end{eqnarray}
where $d\Omega^2 = d\theta^2 + \sin{\theta}^2 d\phi^2$ ($\theta$
and $\phi$ are continuous through the surface).
We define $\Sigma$ in $M^+$ by $r_+ = R(\tau)$ and in $M^-$
as $r_- = R(\tau)$. On $\Sigma$ we choose coordinates
$(\theta, \phi, \tau)$. We seek to determine the equation
governing the evolution of the surface and the stress-energy
of the surface. \\

We demonstrate how to achieve this in the session below. Once again
everything from this point to the end of the subsection is Maple
output. The specification of the surfaces follows exactly as in
the Schwarzschild case in the previous example and so we omit
the input portion of this process in the interests of brevity. \\

\noindent (Prior to this point we defined a surface {\tt sout} in the metric
(\ref{eqn-staticF}) which we
labeled {\tt staticF} and a surface {\tt sint} in (\ref{eqn-staticf}), labeled
{\tt staticf}.)

\noindent We now identify these two surfaces by using {\tt join}.

\begin{mapleinput}
join(sout,sint);
\end{mapleinput}
\begin{maplettyout}
sout and sint are now joined.
The default metric name is sout.
The exterior metric is: sout
The interior metric is: sint

\end{maplettyout}
\begin{maplelatex}
\[
{\it CPU\:Time\:}=.050
\]
\end{maplelatex}
\begin{maplelatex}
\[
{\it For\:the\:sout\:spacetime:}
\]
\end{maplelatex}
\begin{maplelatex}
\[
{\it \:Jump\:from\:defaultMetric\:-\:Mint}
\]
\end{maplelatex}
\begin{maplelatex}
\[
{\it Jump[g(dn,dn),sint]}={\it All\:components\:are\:zero}
\]
\end{maplelatex}
\begin{maplelatex}
\[
{\it staticF}
\]
\end{maplelatex}
\noindent We now calculate $S_i^j$ (the junction package object {\tt
S3(dn,up)}).
Since this is non-zero it is clear that there is a thin shell
seperating {\tt staticF} and {\tt staticf}.

\begin{mapleinput}
grcalc(S3(dn,up));
\end{mapleinput}
\begin{maplettyout}
Created definition for K(dn,up)

\end{maplettyout}
\begin{maplelatex}
\[
{\it CPU\:Time\:}=3.150
\]
\end{maplelatex}
\begin{mapleinput}
gralter(S3(dn,up), factor);
\end{mapleinput}
\begin{maplettyout}
Component Alteration of a grtensor object:

Applying routine factor to object S3(dn,up)

\end{maplettyout}
\begin{maplelatex}
\[
{\it CPU\:Time\:}=.167
\]
\end{maplelatex}
To improve the appearance of the output we make use of GRTensor's ability
to represent derivatives as subscripts (so $dR(\tau)/d\tau \rightarrow R_\tau$)
via the {\tt autoAlias} command. This command resides in the {\tt grtools}
library, which we now load.

\begin{mapleinput}
readlib(grtools):
\end{mapleinput}
Now we apply {\tt autoAlias} via {\tt grmap}.

\begin{mapleinput}
grmap(S3(dn,up), autoAlias, 'x');
\end{mapleinput}
\begin{maplettyout}
Applying routine autoAlias to S3(dn,up)

\end{maplettyout}
\begin{mapleinput}
grdisplay(S3(dn,up));
\end{mapleinput}
\begin{maplelatex}
\[
{\it For\:the\:sout\:spacetime:}
\]
\end{maplelatex}
\begin{maplelatex}
\[
{\it Intrinsic\:stress-energy}
\]
\end{maplelatex}
\begin{maplelatex}
\begin{eqnarray*}
\lefteqn{{{\it S\:}_{{ \theta}}}\,{}^{{ \theta}}= - \,
{\displaystyle \frac {1}{16}} \left( {\vrule
height0.58em width0em depth0.58em} \right. \! \!  - 2\,\sqrt {
{\rm f}(\,{\rm R}(\,{ \tau}\,)\,) + {{R}_{{ \tau}}}^{2}}\,{\rm F}
(\,{\rm R}(\,{ \tau}\,)\,) - 2\,\sqrt {{\rm f}(\,{\rm R}(\,{ \tau
}\,)\,) + {{R}_{{ \tau}}}^{2}}\,{{R}_{{ \tau}}}^{2}} \\
 & & \mbox{} + 2\,\sqrt {{\rm F}(\,{\rm R}(\,{ \tau}\,)\,) + {{R}
_{{ \tau}}}^{2}}\,{\rm f}(\,{\rm R}(\,{ \tau}\,)\,) + 2\,\sqrt {
{\rm F}(\,{\rm R}(\,{ \tau}\,)\,) + {{R}_{{ \tau}}}^{2}}\,{{R}_{{
 \tau}}}^{2} \\
 & & \mbox{} - \sqrt {{\rm f}(\,{\rm R}(\,{ \tau}\,)\,) + {{R}_{{
 \tau}}}^{2}}\,{\rm R}(\,{ \tau}\,)\,{\rm D}(\,{F}\,)(\,{\rm R}(
\,{ \tau}\,)\,) - 2\,\sqrt {{\rm f}(\,{\rm R}(\,{ \tau}\,)\,) + {
{R}_{{ \tau}}}^{2}}\,{\rm R}(\,{ \tau}\,)\,{{R}_{{ \tau}, { \tau}
}} \\
 & & \mbox{} + \sqrt {{\rm F}(\,{\rm R}(\,{ \tau}\,)\,) + {{R}_{{
 \tau}}}^{2}}\,{\rm R}(\,{ \tau}\,)\,{\rm D}(\,{f}\,)(\,{\rm R}(
\,{ \tau}\,)\,) + 2\,\sqrt {{\rm F}(\,{\rm R}(\,{ \tau}\,)\,) + {
{R}_{{ \tau}}}^{2}}\,{\rm R}(\,{ \tau}\,)\,{{R}_{{ \tau}, { \tau}
}} \! \! \left. {\vrule height0.58em width0em depth0.58em}
 \right)  \left/ {\vrule height0.58em width0em depth0.58em}
 \right. \! \!  \left( {\vrule height0.58em width0em depth0.58em}
 \right. \! \! \,{\rm R}(\,{ \tau}\,) \\
 & & \sqrt {{\rm F}(\,{\rm R}(\,{ \tau}\,)\,) + {{R}_{{ \tau}}}^{
2}}\,\sqrt {{\rm f}(\,{\rm R}(\,{ \tau}\,)\,) + {{R}_{{ \tau}}}^{
2}}\,{ \pi}\, \! \! \left. {\vrule
height0.58em width0em depth0.58em} \right)
\end{eqnarray*}
\end{maplelatex}
\begin{maplelatex}
\begin{eqnarray*}
\lefteqn{{{\it S\:}_{{ \phi}}}\,{}^{{ \phi}}= - \,{\displaystyle
\frac {1}{16}} \left( {\vrule height0.58em width0em depth0.58em}
 \right. \! \!  - 2\,\sqrt {{\rm f}(\,{\rm R}(\,{ \tau}\,)\,) + {
{R}_{{ \tau}}}^{2}}\,{\rm F}(\,{\rm R}(\,{ \tau}\,)\,) - 2\,
\sqrt {{\rm f}(\,{\rm R}(\,{ \tau}\,)\,) + {{R}_{{ \tau}}}^{2}}\,
{{R}_{{ \tau}}}^{2}} \\
 & & \mbox{} + 2\,\sqrt {{\rm F}(\,{\rm R}(\,{ \tau}\,)\,) + {{R}
_{{ \tau}}}^{2}}\,{\rm f}(\,{\rm R}(\,{ \tau}\,)\,) + 2\,\sqrt {
{\rm F}(\,{\rm R}(\,{ \tau}\,)\,) + {{R}_{{ \tau}}}^{2}}\,{{R}_{{
 \tau}}}^{2} \\
 & & \mbox{} - \sqrt {{\rm f}(\,{\rm R}(\,{ \tau}\,)\,) + {{R}_{{
 \tau}}}^{2}}\,{\rm R}(\,{ \tau}\,)\,{\rm D}(\,{F}\,)(\,{\rm R}(
\,{ \tau}\,)\,) - 2\,\sqrt {{\rm f}(\,{\rm R}(\,{ \tau}\,)\,) + {
{R}_{{ \tau}}}^{2}}\,{\rm R}(\,{ \tau}\,)\,{{R}_{{ \tau}, { \tau}
}} \\
 & & \mbox{} + \sqrt {{\rm F}(\,{\rm R}(\,{ \tau}\,)\,) + {{R}_{{
 \tau}}}^{2}}\,{\rm R}(\,{ \tau}\,)\,{\rm D}(\,{f}\,)(\,{\rm R}(
\,{ \tau}\,)\,) + 2\,\sqrt {{\rm F}(\,{\rm R}(\,{ \tau}\,)\,) + {
{R}_{{ \tau}}}^{2}}\,{\rm R}(\,{ \tau}\,)\,{{R}_{{ \tau}, { \tau}
}} \! \! \left. {\vrule height0.58em width0em depth0.58em}
 \right)  \left/ {\vrule height0.58em width0em depth0.58em}
 \right. \! \!  \left( {\vrule height0.58em width0em depth0.58em}
 \right. \! \! \,{\rm R}(\,{ \tau}\,) \\
 & & \sqrt {{\rm F}(\,{\rm R}(\,{ \tau}\,)\,) + {{R}_{{ \tau}}}^{
2}}\,\sqrt {{\rm f}(\,{\rm R}(\,{ \tau}\,)\,) + {{R}_{{ \tau}}}^{
2}}\,{ \pi}\, \! \! \left. {\vrule
height0.58em width0em depth0.58em} \right)
\end{eqnarray*}
\end{maplelatex}
\begin{maplelatex}
\begin{eqnarray*}
\lefteqn{{{\it S\:}_{{ \tau}}}\,{}^{{ \tau}}= - \,{\displaystyle
\frac {1}{4}} \left( {\vrule height0.58em width0em depth0.58em}
 \right. \! \!  - \sqrt {{\rm f}(\,{\rm R}(\,{ \tau}\,)\,) + {{R}
_{{ \tau}}}^{2}}\,{\rm F}(\,{\rm R}(\,{ \tau}\,)\,) - \sqrt {
{\rm f}(\,{\rm R}(\,{ \tau}\,)\,) + {{R}_{{ \tau}}}^{2}}\,{{R}_{{
 \tau}}}^{2} + \sqrt {{\rm F}(\,{\rm R}(\,{ \tau}\,)\,) + {{R}_{{
 \tau}}}^{2}}\,{\rm f}(\,{\rm R}(\,{ \tau}\,)\,)} \\
 & & \mbox{} + \sqrt {{\rm F}(\,{\rm R}(\,{ \tau}\,)\,) + {{R}_{{
 \tau}}}^{2}}\,{{R}_{{ \tau}}}^{2} \! \! \left. {\vrule
height0.58em width0em depth0.58em} \right)  \left/ {\vrule
height0.58em width0em depth0.58em} \right. \! \!  \left( \! \,
{\rm R}(\,{ \tau}\,)\,\sqrt {{\rm F}(\,{\rm R}(\,{ \tau}\,)\,) +
{{R}_{{ \tau}}}^{2}}\,\sqrt {{\rm f}(\,{\rm R}(\,{ \tau}\,)\,) +
{{R}_{{ \tau}}}^{2}}\,{ \pi}\, \!  \right) \mbox{\hspace{52pt}}
\end{eqnarray*}
\end{maplelatex}
\noindent Note that since $f(r)$ and $F(r)$ are unspecified their derivatives
cannot be evaluated.
This results in the use of the Maple {\tt D} derivative. We will demonstrate
how to specify specific functions below.

\noindent Next we consider the equation for the history of the shell {\tt
HGeqn}
and the first integral of this equation {\tt evInt1}
(which in the case of spherical symmetry stems from an identity).

\begin{mapleinput}
grcalc(HGeqn, evInt1);
\end{mapleinput}
\begin{maplettyout}
Created definition for n(up)
Created definition for R(dn,dn,up,up)

\end{maplettyout}
\begin{maplelatex}
\[
{\it CPU\:Time\:}=4.767
\]
\end{maplelatex}
\noindent Before displaying the results we factor the expressions.

\begin{mapleinput}
gralter(_,factor);
\end{mapleinput}
\begin{maplettyout}
Component Alteration of a grtensor object:

Applying routine factor to object HGeqn
Applying routine factor to object evInt1

\end{maplettyout}
\begin{maplelatex}
\[
{\it CPU\:Time\:}=.317
\]
\end{maplelatex}
\begin{mapleinput}
grdisplay(_);
\end{mapleinput}
\begin{maplelatex}
\[
{\it For\:the\:sout\:spacetime:}
\]
\end{maplelatex}
\begin{maplelatex}
\[
{\it History\:equation}
\]
\end{maplelatex}
\begin{maplelatex}
\begin{eqnarray*}
\lefteqn{{\it HGeqn\:}= \left( {\vrule
height0.87em width0em depth0.87em} \right. \! \! {\displaystyle
\frac {1}{4}} \left( {\vrule height0.58em width0em depth0.58em}
 \right. \! \!  - {\rm R}(\,{ \tau}\,)\,{ \sigma}(\,{ \theta}, {
\phi}, { \tau}\,)\,\sqrt {{\rm f}(\,{\rm R}(\,{ \tau}\,)\,) + {{R
}_{{ \tau}}}^{2}}\,{\rm D}(\,{F}\,)(\,{\rm R}(\,{ \tau}\,)\,)} \\
 & & \mbox{} - 2\,{\rm R}(\,{ \tau}\,)\,{ \sigma}(\,{ \theta}, {
\phi}, { \tau}\,)\,\sqrt {{\rm f}(\,{\rm R}(\,{ \tau}\,)\,) + {{R
}_{{ \tau}}}^{2}}\,{{R}_{{ \tau}, { \tau}}} - 2\,{\rm R}(\,{ \tau
}\,)\,{ \sigma}(\,{ \theta}, { \phi}, { \tau}\,)\,\sqrt {{\rm F}(
\,{\rm R}(\,{ \tau}\,)\,) + {{R}_{{ \tau}}}^{2}}\,{{R}_{{ \tau},
{ \tau}}} \\
 & & \mbox{} - {\rm R}(\,{ \tau}\,)\,{ \sigma}(\,{ \theta}, {
\phi}, { \tau}\,)\,\sqrt {{\rm F}(\,{\rm R}(\,{ \tau}\,)\,) + {{R
}_{{ \tau}}}^{2}}\,{\rm D}(\,{f}\,)(\,{\rm R}(\,{ \tau}\,)\,) \\
 & & \mbox{} + 4\,{\rm P}(\,{ \theta}, { \phi}, { \tau}\,)\,
\sqrt {{\rm f}(\,{\rm R}(\,{ \tau}\,)\,) + {{R}_{{ \tau}}}^{2}}\,
{\rm F}(\,{\rm R}(\,{ \tau}\,)\,) + 4\,{\rm P}(\,{ \theta}, {
\phi}, { \tau}\,)\,\sqrt {{\rm f}(\,{\rm R}(\,{ \tau}\,)\,) + {{R
}_{{ \tau}}}^{2}}\,{{R}_{{ \tau}}}^{2} \\
 & & \mbox{} + 4\,{\rm P}(\,{ \theta}, { \phi}, { \tau}\,)\,
\sqrt {{\rm F}(\,{\rm R}(\,{ \tau}\,)\,) + {{R}_{{ \tau}}}^{2}}\,
{\rm f}(\,{\rm R}(\,{ \tau}\,)\,) + 4\,{\rm P}(\,{ \theta}, {
\phi}, { \tau}\,)\,\sqrt {{\rm F}(\,{\rm R}(\,{ \tau}\,)\,) + {{R
}_{{ \tau}}}^{2}}\,{{R}_{{ \tau}}}^{2} \! \! \left. {\vrule
height0.58em width0em depth0.58em} \right)  \left/ {\vrule
height0.58em width0em depth0.58em} \right. \! \!  \left( {\vrule
height0.58em width0em depth0.58em} \right. \! \!  \\
 & & {\rm R}(\,{ \tau}\,)\,\sqrt {{\rm F}(\,{\rm R}(\,{ \tau}\,)
\,) + {{R}_{{ \tau}}}^{2}}\,\sqrt {{\rm f}(\,{\rm R}(\,{ \tau}\,)
\,) + {{R}_{{ \tau}}}^{2}}\, \! \! \left. {\vrule
height0.58em width0em depth0.58em} \right) = \\
 & &  - \,{\displaystyle \frac {{\rm R}(\,{ \tau}\,)\,{\rm D}(\,{
F}\,)(\,{\rm R}(\,{ \tau}\,)\,) + {\rm F}(\,{\rm R}(\,{ \tau}\,)
\,) - {\rm R}(\,{ \tau}\,)\,{\rm D}(\,{f}\,)(\,{\rm R}(\,{ \tau}
\,)\,) - {\rm f}(\,{\rm R}(\,{ \tau}\,)\,)}{{\rm R}(\,{ \tau}\,)
^{2}}} \! \! \left. {\vrule height0.87em width0em depth0.87em}
 \right)
\end{eqnarray*}
\end{maplelatex}
\begin{maplelatex}
\[
{}
\]
\end{maplelatex}
\begin{maplelatex}
\begin{eqnarray*}
\lefteqn{{\it evInt1\:}= \left( {\vrule
height0.79em width0em depth0.79em} \right. \! \! {{R}_{{ \tau}}}
^{2}={\displaystyle \frac {1}{64}} \left( {\vrule
height0.44em width0em depth0.44em} \right. \! \!  - 64\,{ \pi}^{2
}\,{ \sigma}(\,{ \theta}, { \phi}, { \tau}\,)^{2}\,{\rm R}(\,{
\tau}\,)^{4}\,{r}^{2} + {r}^{4}\,{\rm F}(\,{r}\,)^{2} - 2\,{r}^{4
}\,{\rm F}(\,{r}\,)\,{\rm f}(\,{r}\,) + {r}^{4}\,{\rm f}(\,{r}\,)
^{2}} \\
 & & \mbox{} + 64\,(\,{r}^{2}\,)^{3/2}\,{ \pi}^{2}\,{ \sigma}(\,{
 \theta}, { \phi}, { \tau}\,)^{2}\,{\rm R}(\,{ \tau}\,)^{3} - 32
\,(\,{r}^{2}\,)^{3/2}\,{ \pi}^{2}\,{ \sigma}(\,{ \theta}, { \phi}
, { \tau}\,)^{2}\,{\rm R}(\,{ \tau}\,)^{3}\,{\rm F}(\,{r}\,) \\
 & & \mbox{} - 32\,(\,{r}^{2}\,)^{3/2}\,{ \pi}^{2}\,{ \sigma}(\,{
 \theta}, { \phi}, { \tau}\,)^{2}\,{\rm R}(\,{ \tau}\,)^{3}\,
{\rm f}(\,{r}\,) + 256\,{ \pi}^{4}\,{ \sigma}(\,{ \theta}, { \phi
}, { \tau}\,)^{4}\,{\rm R}(\,{ \tau}\,)^{6}\,{r}^{2} \!
\! \left. {\vrule height0.44em width0em depth0.44em} \right)
 \left/ {\vrule height0.44em width0em depth0.44em} \right. \! \!
 \left( {\vrule height0.44em width0em depth0.44em} \right. \! \!
\,{ \pi}^{2} \\
 & & { \sigma}(\,{ \theta}, { \phi}, { \tau}\,)^{2}\,{\rm R}(\,{
\tau}\,)^{4}\,{r}^{2}\, \! \! \left. {\vrule
height0.44em width0em depth0.44em} \right)  \! \! \left. {\vrule
height0.79em width0em depth0.79em} \right) \mbox{\hspace{264pt}}
\end{eqnarray*}
\end{maplelatex}
\noindent Now we define functions for $f$ and $F$. We consider the Israel
thin-shell example and
hence define $f(r) = 1$ and $F(r) = 1 - 2m/r$ so we have a Minkowski interior
and a
Schwarzschild exterior. These functions are defined as Maple procedures. (This
ensures that the defered
derivatives will evaluate. Note that a simple substitution of e.g. $f(R(t)) =
1-2m/R(t)$ would not affect
\verb|D(f)(R(t))| since \verb|f(R(t))| does not appear explicitly).
A Maple procedure is declared in a statement of the form:
\verb|procedureName := proc(arguments) procedure_body end:|

\begin{mapleinput}
f := proc(r) RETURN(1); end:
\end{mapleinput}
\begin{mapleinput}
F := proc(r) RETURN(1-2*m/r); end:
\end{mapleinput}
\noindent To repeat Israel's dust shell analysis we must set the pressure to
zero, which we now do.

\begin{mapleinput}
grmap(_, subs, P(theta,phi,tau)=0,'x');
\end{mapleinput}
\begin{maplettyout}
Applying routine subs to HGeqn
Applying routine subs to evInt1

\end{maplettyout}
\begin{mapleinput}
gralter(_,power,expand);
\end{mapleinput}
\begin{maplettyout}
Component Alteration of a grtensor object:

Applying routine simplify[power] to object HGeqn
Applying routine simplify[power] to object evInt1
Applying routine expand to object HGeqn
Applying routine expand to object evInt1

\end{maplettyout}
\begin{maplelatex}
\[
{\it CPU\:Time\:}=1.116
\]
\end{maplelatex}
\begin{mapleinput}
grdisplay(_);
\end{mapleinput}
\begin{maplelatex}
\[
{\it For\:the\:sout\:spacetime:}
\]
\end{maplelatex}
\begin{maplelatex}
\[
{\it History\:equation}
\]
\end{maplelatex}
\begin{maplelatex}
\begin{eqnarray*}
\lefteqn{{\it HGeqn\:}=} \\
 & &  \left( \! \, - \,{\displaystyle \frac {1}{2}}\,
{\displaystyle \frac {{ \sigma}(\,{ \theta}, { \phi}, { \tau}\,)
\,{m}}{{\rm R}(\,{ \tau}\,)^{2}\,\sqrt {1 - 2\,{\displaystyle
\frac {{m}}{{\rm R}(\,{ \tau}\,)}} + {{R}_{{ \tau}}}^{2}}}} -
{\displaystyle \frac {1}{2}}\,{\displaystyle \frac {{ \sigma}(\,{
 \theta}, { \phi}, { \tau}\,)\,{{R}_{{ \tau}, { \tau}}}}{\sqrt {1
 - 2\,{\displaystyle \frac {{m}}{{\rm R}(\,{ \tau}\,)}} + {{R}_{{
 \tau}}}^{2}}}} - {\displaystyle \frac {1}{2}}\,{\displaystyle
\frac {{ \sigma}(\,{ \theta}, { \phi}, { \tau}\,)\,{{R}_{{ \tau}
, { \tau}}}}{\sqrt {1 + {{R}_{{ \tau}}}^{2}}}}=0\, \!  \right)
\end{eqnarray*}
\end{maplelatex}
\begin{maplelatex}
\[
{}
\]
\end{maplelatex}
\begin{maplelatex}
\[
{\it evInt1\:}= \left( \! \,{{R}_{{ \tau}}}^{2}= - 1 +
{\displaystyle \frac {1}{16}}\,{\displaystyle \frac {{m}^{2}}{{
\pi}^{2}\,{ \sigma}(\,{ \theta}, { \phi}, { \tau}\,)^{2}\,{\rm R}
(\,{ \tau}\,)^{4}}} + {\displaystyle \frac {(\,{r}^{2}\,)^{3/2}\,
{m}}{{r}^{3}\,{\rm R}(\,{ \tau}\,)}} + 4\,{ \pi}^{2}\,{ \sigma}(
\,{ \theta}, { \phi}, { \tau}\,)^{2}\,{\rm R}(\,{ \tau}\,)^{2}\,
 \!  \right)
\]
\end{maplelatex}
\noindent The above expressions correspond to the results given in
\cite{Israel66}. Note that we did not have
to integrate to obtain the first integral of the evolution equation.

\subsection{A complicated Minkowski junction}
\label{sec-toy}
In this section we describe the junction of the Minkowski metric in
the form
\begin{equation}
\label{eqn-toy}
ds_-^2 = \frac{r^2+u^2}{r^2+a^2} dr^2 +
 \frac{r^2+u^2}{a^2-u^2} du^2+
 \frac{(a^2-u^2)(r^2+a^2)}{a^2} d\phi^2 - dt^2
\end{equation}
to a metric $ds_+^2$ of the same form, but with coordinates
$(R, U,\Phi,T)$ and parameter $A$. These metrics arise
from setting $m=0$ in a Kerr metric and choosing $u= a~ \cos{\theta})$.
The coordinate names $r$ and $R$ while standard
are quite misleading since surfaces of constant ``radius'' describe
spheroids with oblateness governed by $a$ or $A$. We seek to join
a surface of constant ``radius'' in $M^-$ to $M^+$. The surface
in $M^+$ will be some function of $R$ and $U$.
This example demonstrates
{\tt GRJunction}'s ability to handle non-spherical
surfaces in a case where we know a boundary surface must result. \\

The transformations from e.g. (\ref{eqn-toy}) to the Minkowski metric
in spherical form (with coordinates $\tilde r$ and $\tilde \theta$)
are given by
\begin{eqnarray}
\label{eqn-toyxform}
u & = & a~\cos{\theta} \\
\tilde r^2 & = & r^2 + a^2 \sin{\theta}^2 \nonumber \\
\tilde r cos{\tilde \theta} & = & r \cos{\theta}.
\end{eqnarray}
If we take the definition of $\Sigma^-$ as $r=X$ ($X$ a constant) then
we can use (\ref{eqn-toyxform}) to determine that the
the definition of the corresponding surface in $M^+$ has parametric
equation:
\begin{eqnarray}
0 & = & X - (R^2+A^2-U^2-a^2+\frac{1}{4}(-2R^2A^2+2a^2A^2-2A^4+2A^2U^2+
   \nonumber \\
 & & 2(R^4A^4-2R^2A^4a^2 + 2R^2A^6 - 2R^2A^4U^2+a^4A^4-2a^2A^6 +
   \nonumber \\
 & & 2a^2A^4U^2 + A^8 - 2A^6U^2 + A^4U^4 + 4R^2U^2a^2A^2)^{(1/2)})/A^2)^{(1/2)}
\end{eqnarray}
and the relations $x^\alpha_+ = x^\alpha_+(\xi^i)$ are
\begin{eqnarray}
R & = & (-A^2 + \frac{1}{2a^2}(-X^2a^2+a^2A^2+a^2u^2-a^4+(X^4a^4-2X^2a^4A^2
\nonumber \\
 & &  - 2X^2a^4u^2 + 2X^2a^6 + a^4A^4 + 2a^4A^2u^2 - 2a^6A^2+a^4u^4-2a^6u^2+a^8
\nonumber \\
 & & +4X^2u^2A^2a^2)^{1/2}) + X^2+a^2-u^2)^{1/2} \\
U & = & \frac{1}{a\sqrt{2}}(-X^2a^2+a^2A^2+a^2u^2-a^4+
 (X^4a^4-2X^2a^4A^2-2X^2a^4u^2  \nonumber \\
 & & + 2X^2a^6 +a^4A^4 + 2a^4A^2u^2-2a^6 A^2 + a^4u^4 \nonumber \\
 & & - 2a^6u^2 + a^8 + 4X^2u^2A^2a^2)^{1/2})^{1/2} \\
\Phi & = & \phi \\
t & = & T_+(\tau).
\end{eqnarray}
In this case it not clear which sign we should choose for the
normal vector in $M^+$.
However since we know {\em a priori} that a boundary surface
must result if the package does not reach this result then we can
consider the other choice of normal sign for $n^+_\alpha$.
(In this case we choose the plus sign in (\ref{eqn-ndef})). \\

The package determines $\left[g_{ij}\right]=0$
and that $K^-_{i j}$ is:
\begin{eqnarray}
\label{eqn-Kmink}
K^-_{u u} & = & \frac{X \sqrt{X^2+a^2}}{(a^2-u^2)\sqrt{X^2+u^2}} \nonumber \\
K^-_{\phi \phi} & = &  \frac{X \sqrt{X^2+a^2} (a^2-u^2)}{a^2\sqrt{X^2+u^2}}.
\end{eqnarray}
With careful direction of the computer simplification
the junction package determines that
$K^+_{ij}$ is also given by (\ref{eqn-Kmink}).
\footnote{In an earlier attempt to determine $K^+_{ij}$
(using Boyer-Lindquist coordinates)
we encountered the Maple error ``Object too Large'' which occurs on
32-bit machines
when an expression contains more than 64,000 terms. On present
day workstations this limit can be reached in a matter of minutes
and often little can be done to circumvent this Maple limitation.} \\

\subsection{Joining Kerr to Kerr}
In this section we consider joining the
``interior'' region of one Kerr spacetime $\cal M^-$ in
Boyer-Lindquist form
\begin{eqnarray}
ds^2 & = & \rho \left(\frac{dr_-^2}{\delta} + d\theta_-^2 \right)
 + (r_-^2 + a^2) \sin{\theta_-}^2 d\phi_-^2 - dt_-^2 \nonumber \\
 & & + \frac{2mr_-}{\rho} (a \sin{\theta_-}^2 d\phi_- - dt_-)^2 \nonumber \\
 & & \rho = r_-^2 + a^2 \cos{\theta_-}^2,~~\delta = r_-^2 - 2mr_- + a^2
\end{eqnarray}
with mass $m$ and angular momentum $a$ to the ``exterior'' region
of another Kerr spacetime $\cal M^+$
with mass $M$ and angular momentum $A$
(we use coordinates $(r_\pm, \theta_\pm, \phi_\pm, t_\pm)$ for
$\cal M^\pm$). Such a problem would arise
if a thin shell of matter was constructed around a Kerr
black hole (i.e. a Dyson sphere). The general problem is
an extremely difficult one as the ``toy'' problem
$m=M=0$ indicates (see section \ref{sec-toy}). In this
section we limit the discussion to an expansion to first order in $a$ and $A$,
since
to this order the surface in $\cal M^\pm$ is spherical. \\

In the treatment of timelike spherical surfaces it is customary
to use transformations such as $r=R(\tau)$ and $t=T(\tau)$ and
then use $u_\alpha u^\alpha = -1$ to eliminate
$\partial T(\tau) / \partial \tau$. Note that in these spherical
cases this produces $g_{\tau \tau} = -1$ on $\Sigma$ as desired.
In all spherical cases $T$ will be strictly a function of
$\tau$ but this fails to be true in Kerr.
For Kerr spacetimes to achieve $g_{\tau \tau} = -1$ we can use
the same idea but if we blindly use $t=T(\tau)$ we discover
that in actuality $t=T(\tau, \tilde \theta)$. We can try again - using
$\tilde \theta$ as an argument and we do get tau as the proper time
on the shell but now
a $\partial T(\tau,\tilde \theta) / \partial \tilde \theta$ appears in
$g_{\tilde \theta \tilde \theta}, g_{\tilde \theta \tau}$
and $g_{\tilde \theta \tilde \phi}$ .
Since we know $\partial T / \partial \tau$ is
merely a function of $\tilde \theta$ we can integrate trivially but
this makes those components with a $\partial T / \partial \theta$
depend linearly on $\tau$ and the metric components are explicitly
dependent on proper time. Hence forcing $\tau$ to be the proper time
on the surface comes at considerable expense. Fortunately these problems
do not arise in the order $(a,A)$ expansions. \\

To facilitate the matching of the $g_{ij}$ on $\Sigma$ we
eliminate the $g_{\tilde \phi \tau}$ terms
on $\Sigma$ by using a transformation to the zero angular momentum (ZAMO)
frame
in the definition of the surface. The transformations are
\begin{equation}
r_\pm=R, \theta_\pm = \tilde \theta,
\phi_\pm = \tilde \phi - \Omega_\pm~T_\pm(\tau, \theta),
t_\pm = T_\pm(\tau, \theta)
\end{equation}
where $\Omega \equiv g_{\phi t} / g_{\phi \phi}$. Using {\tt GRjunction}
we calculate the metric and second fundamental form on the surface
and only then expand to order $a$, identify $\Sigma^\pm$ and
calculate $S_{ij}$. \\

The package first determines
\begin{equation}
ds^2 \mid_{\Sigma^\pm} = R^2 d\tilde \theta^2 +
R^2 \sin{\tilde \theta}^2 d\tilde \phi^2 - d\tau^2
\end{equation}
and consequently $\left[ g_{ij} \right] = 0$. \\

The resulting stress-energy of the shell is (from this point on
we drop the tildes on $\theta$ and $\phi$):
\begin{eqnarray*}
S_\theta^\theta & = & \frac{ f(R-M) - F(R-m)}{8 \pi R^2 f F} \\
S_\phi^\phi & = & S_\theta^\theta \\
S_\phi^\tau & = & R^2 sin^2 \theta~S_\tau^\phi \\
S_\tau^\phi & = & \frac{3 (am-AM)}{8 \pi R^2} \\
S_\tau^\tau & = & \frac{f(R-2M) - F(R-2m)}{ 4 \pi R^2 f F}
\end{eqnarray*}
where $f = \sqrt{1-2m/R}, F= \sqrt{1-2M/R}$. \\

Note the appearance of mixed term $S_\phi^\tau$ which precludes
a standard phenomenological interpretation (i.e. $\sigma$
as given by (\ref{eqn-sigma}) is not
an eigenvector of $S_{ij}$).
To first order
we have
\begin{equation}
\Omega_- =  - \frac{a m}{R^3},~~~\Omega_+ = - \frac{ A M}{R^3}
\end{equation}
and if we require $\Omega_- = \Omega_+$ then the mixed terms
vanish. This now permits standard phenomenological interpretation
of the shell and we can now interpret the density and pressures in the
usual way. Note that in this case
the pressure is isotropic and hoop stresses do not arise. \\

We can easily extend this calculation to the dynamic case with
$R \rightarrow R(\tau)$ in the above. The resulting stress
energy tensor is:
\begin{eqnarray*}
S_\theta^\theta & = &
    \frac{ R^2 \ddot{R} (f-F) + f(R-M) - F(R-m)}{8 \pi R^2 f F} \\
S_\phi^\phi & = & S_\theta^\theta \\
S_\phi^\tau & = & R^2 sin^2 \theta~S_\tau^\phi \\
S_\tau^\phi & = & \frac{3}{8 \pi}
    \frac{A M f (R^5 (\ddot{R}^2 R + R -2M))^{1/2} -
          a m F (R^5 (\ddot{R}^2 R + R -2m))^{1/2} }{ R^7 f F} \\
S_\tau^\tau & = & \frac{f(R-2M) - F(R-2m)}{ 4 \pi R^2 f F}
\end{eqnarray*}
where $\dot{} = d/d\tau$.

\section{Verification}
In developing the junction package we have re-executed a number
of calculations performed in the literature to verify that the
package can reproduce these results. We list some of the
verifications we have performed in Table 5
(Maple worksheets for these verifications are available by
ftp \cite{ftpSite}.) \\

\begin{table}
\begin{center}
\begin{tabular}{|l|l|l|} \hline
{\bf Problem} &  {\bf Verified} & {\bf Comments} \\ \hline
Dust shell \cite{Israel66} &  $K_{ij}$  & Minkowski-Shell-Schw.\\
     & evolution equation & dynamics \\
 & & \\
spherical & $K_{ij}$ & Checked form of $K_{ij}$ for general \\
symmetry \cite{Berezin}  &          & spherical symmetry  \\
 & & \\
inhomogeneous &   $\left[ g_{ij} \right] = 0$ & planar symmetry \\
slab cosmology \cite{LakeSlab} & $\left[ K_{ij} \right] = 0$ &  \\
 & & \\
Schwarzschild &  $S_{ij}$, $\sigma$, $p$ & ``Thin-shell'' wormhole \\
wormhole \cite{Visser} &                        &  formed by joining Schw.
exterior \\
 & & to Schw. exterior \\
 & & \\

Schw- deSitter &  $S_{ij}$   & Spherical symmetry but uses \\
shell \cite{Frolov}   &   & coordinates $(r,v,\theta,\phi)$ \\
 & & \\
Rotating dust & $\left[ g_{ij} \right] =0 , K^+_{ij}$ to order $a^3$ &
   match to Kerr exterior \\
shell \cite{DeLaCruz} & $S_{ij}$ to order $a$  &  in small $a$ limit \\
 & & (non-spherical surface) \\
 & & \\

collapsing, rotating & $S_{ij}$ & Match to Kerr to order $a$ \\
dust shell \cite{Lindbloom} &  & (dynamic shell) \\ \hline
\end{tabular}
\end{center}
\caption{Results verified with {\tt GRjunction}}
\end{table}

In addition we have verified the identities (\ref{eqn-claw}) and
(\ref{eqn-elaw})
for a variety of spacetimes including a number of non-spherical wormhole
spacetimes. \\


\newpage

\begin{thebibliography}{99}

\bibitem{Darmois} G. Darmois (1927) {\em M\'{e}morial de Sciences
Math\'{e}matiques, Fascicule XXV}, ``Les equations de la
gravitation einsteinienne'', Chapitre V.

\bibitem{Israel66} W. Israel (1966) Thin shells in general relativity.
Il. Nuovo Cim. 66:1.

\bibitem{OppSnyder} J.R. Oppenheimer and H. Snyder (1939)
Phys. Rev. {\bf 56}:455.

\bibitem{ftpSite} The package can be obtained by anonymous
ftp from astro.queensu.ca (103.15.26.39) in the directory
\verb|/pub/grtensor| or from the
web site \verb|http://astro.queensu.ca/~grtensor/GRHome.html|.

\bibitem{Maple} MapleV Release 3. Waterloo Maple Software. Waterloo,
Ontario, Canada.

\bibitem{GRT} P. Musgrave, D. Pollney and K. Lake (1995)
GRTensorII software. Available as indicated in \cite{ftpSite}.

\bibitem{MTW} C.W. Misner, K.S. Thorne and J.A. Wheeler (1973) {\em
Gravitation}.
W.H. Freeman and Company.

\bibitem{LakeBrazil} K. Lake (1988) In: {\em
Fifth Brazialian School of Cosmology and Gravitation} Ed:
M. Novello. 1-82.

\bibitem{Berezin} V.A. Berezin, V.A. Kuzmin and I.I. Tkachev
(1987) Phys. Rev. D {\bf 36}: 2919.

\bibitem{juncRevLast} C. Barrab\`{e}s and W. Israel. (1991)
Phys.~Rev.~D {\bf 43}: 1129.

\bibitem{LakeIdentity} K. Lake (1979) Phys. Rev. D {\bf 19}:2847.

\bibitem{Sato} H. Sato (1986) Prog. Theor. Phys. {\bf 76}:1250.

\bibitem{Maeda} N. Sakai and K. Maeda (1994) Phys. Rev. D {\bf 50}:5425.

\bibitem{Lich} A. Lichnerowicz (1955) {\em Th\'{e}ories Relativistes
de la Gravitation et de l'Electromagn\'{e}tisme}
Masson, Paris. Chapitre III.

\bibitem{Redmount} I.H. Redmount (1985) Prog. Theor. Phys. {\bf 73}: 1401.

\bibitem{LakeSlab} K. Lake (1992) Ap. J. {\bf 401}:L1.

\bibitem{Visser} E. Poisson and M. Visser (1995) gr-qc 9506083. To appear
in Phys. Rev. D.

\bibitem{Frolov} V.P. Frolov, M.A. Markov and V.F. Mukhanov (1990)
Phys. Rev. D {\bf 41}:383.

\bibitem{DeLaCruz} V. De La Cruz and W. Israel (1967) Phys. Rev. {\bf
170}:1187.

\bibitem{Lindbloom} L. Lindblom and D. R. Brill (1974) Phys. Rev. D {\bf
10}:3151.


\end{thebibliography}
\end{document}